\documentclass[11pt]{article}
\usepackage{graphicx}
\usepackage{epsfig}
\usepackage{hyperref}
\usepackage[usenames,dvipsnames]{color}
\usepackage{wrapfig}
\usepackage{amsmath}
\def\lsim{\raise0.3ex\hbox{$<$\kern-0.75em\raise-1.1ex\hbox{$\sim$}}}
\def\gsim{\raise0.3ex\hbox{$>$\kern-0.75em\raise-1.1ex\hbox{$\sim$}}}

\def\mean#1{\left<#1\right>}
\def\Journal#1#2#3#4{{#1}{\bf #2} (#4) #3}

\def\EPJC{{Eur. Phys. J. C}}
\def\JPG{{J. Phys. G}}

\def\NIMA{{Nucl. Instrum. Methods A}}
\def\NPA{{Nucl. Phys. A}}

\def\PLB{{Phys. Lett. B}}
\def\PLC{Phys. Repts.\ }

\def\PRL{Phys. Rev. Lett.\ }
\def\PRD{{Phys. Rev. D}}
\def\PRC{{Phys. Rev. C}}

\def\ARNPS{{Ann. Rev. Nucl. Part. Sci.\ }}

\def\QGP{{\color{Red} Q}{\color{Blue} G}{\color{Green} P}} 
\def\QCD{{\color{Red} Q}{\color{Green} C}{\color{Blue} D}} 

\normalsize
\begin{document}
\title{Highlights from BNL-RHIC}
\author{M.~J.~Tannenbaum 
\thanks{Research supported by U.S. Department of Energy, DE-AC02-98CH10886.}
\\ Physics Department, 510c,\\
Brookhaven National Laboratory,\\
Upton, NY 11973-5000, USA\\
mjt@bnl.gov}\maketitle
\maketitle
\thispagestyle{empty}

\section{Introduction}\label{sec:introduction}
Since the AGS and SpS fixed target heavy-ion programs began operations in 1986, BNL and CERN have been the principal laboratories for measurements of high energy nucleus-nucleus collisions which create nuclear matter in conditions of extreme temperature and density~\cite{EriceProcPR,JanMikeBook}. The kinetic energy of the incident projectiles is dissipated in the large volume of nuclear matter involved in the reaction.  At large energy or baryon densities, a phase transition is expected from a state of nucleons containing confined quarks and gluons to a state of ``deconfined'' (from their individual nucleons) quarks and gluons, in chemical and thermal equilibrium, covering a volume that is many units of the confining length scale. This state of nuclear matter was originally given the name Quark Gluon Plasma (\QGP)~\cite{Shuryak80}, a plasma being an ionized gas. 

The startup of the Relativistic Heavy Ion Collider (RHIC) at BNL in the year 2000 provided a major advance in the field, leading to the discovery of the \QGP\ at RHIC, which was announced on April 19, 2005.   
The results at RHIC~\cite{EriceProcPR} indicated that instead of behaving like a gas of free quarks and gluons, the matter created in heavy ion collisions at nucleon-nucleon c.m. energy $\sqrt{s_{NN}}=200$ GeV appears to be more like a {\em liquid}. This matter interacts much more strongly than originally expected, as elaborated in peer reviewed articles by the 4 RHIC experiments~\cite{BRWP,PHWP,STWP,PXWP}, which inspired the theorists~\cite{THWPS} to give it the new name ``s\QGP" (strongly interacting \QGP). These properties were quite different from the properties of the CERN SpS fixed-target heavy ion results which led to the claim of ``new state of matter'', in a press-conference~\cite{CERNBaloney} on February 10, 2000,  that was neither peer-reviewed nor published. However, results in the past two years from Pb+Pb measurements at the CERN-LHC at $\sqrt{s_{NN}}=2760$ GeV (2.76 TeV) confirm the RHIC discoveries~\cite{EriceProcPR,JanMikeBook} and add some new information---notably with fully reconstructed jets~\cite{ATLASdijet,CMSdijet}. 

$J/\Psi$ suppression, proposed by Matsui and Satz in 1986~\cite{MatsuiSatz86} as the `gold-plated' signature for deconfinement, suffered from the significant $J/\Psi$ suppression observed in the Cold Nuclear Matter (CNM) of p+A collisions~\cite{E772} ($A^{0.92}$) relative to the point-like scaling ($A^{1.0})$ from p-p collisions expected for the large mass scale. Although, ``anomalous'' $J/\Psi$ suppression, i.e. more than the estimated CNM effect, was discovered at the CERN SPS heavy ion program and is its main claim to fame, the later development of $J/\Psi$ measurements has not been concerned with $J/\Psi$ suppression as a signature of deconfinement, but rather with the strong c.m. energy dependence of the CNM effect and the possibility of regeneration of $J/\Psi$ from recombination of the large number of $c$ and $\bar c$ quarks produced in the \QGP\ (e.g. see ref.~\cite{EriceProcPR}).  
Nevertheless, the search for $J/\Psi$ suppression (as well as thermal photon/dilepton radiation from the \QGP ) drove the design of the RHIC experiments~\cite{RHICNIM} and the ALICE experiment at the LHC~\cite{LHCJINST}. 

\section{The RHIC machine in 2012}
\begin{figure}[!h]
\begin{center}
\includegraphics[width=0.90\textwidth]{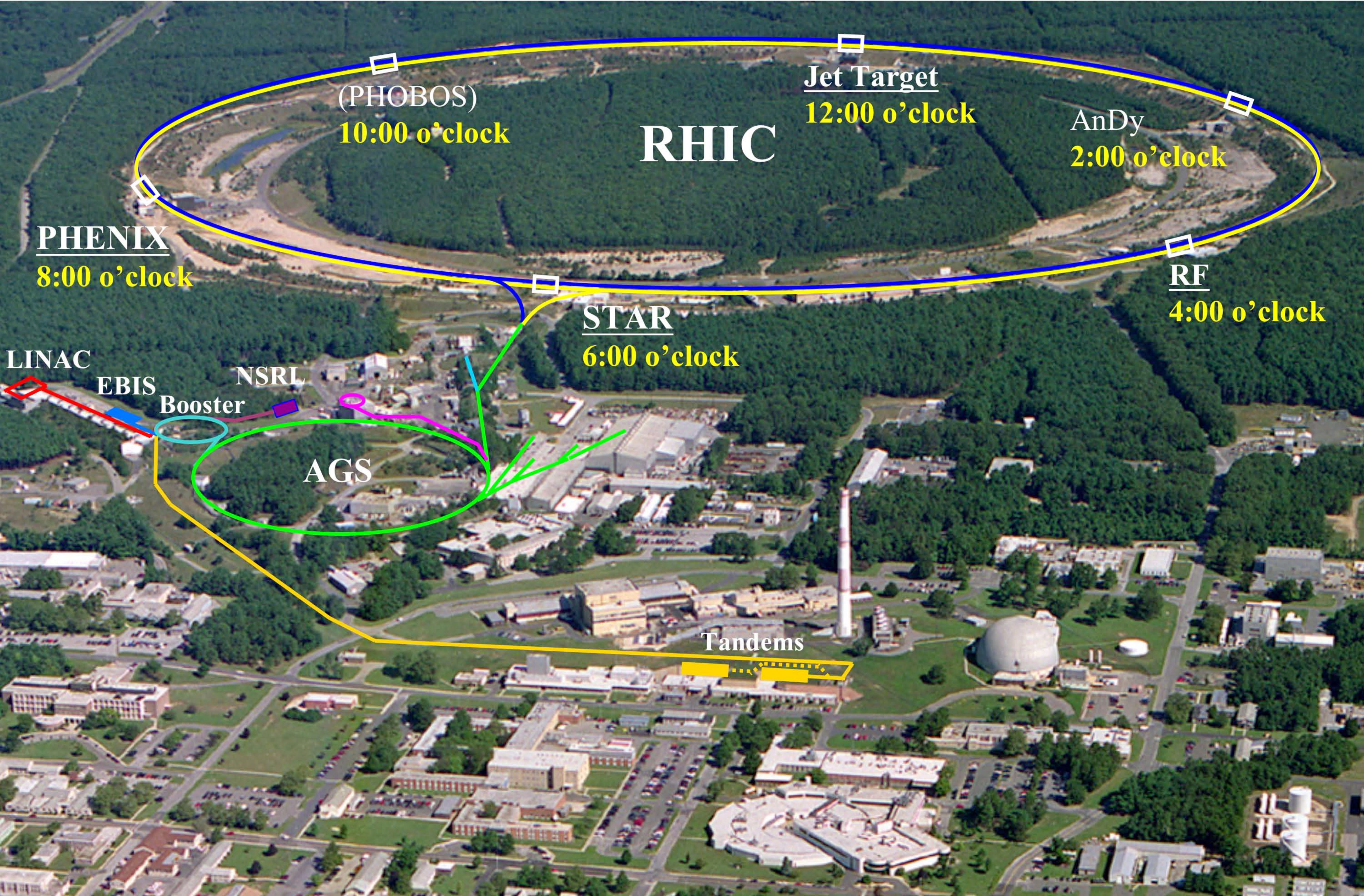}
\end{center}\vspace*{-1.5pc}
\caption[]{Aerial view of the RHIC facility~\cite{RHICNIM}. The six crossing points are labelled as on a clock. The two principal experiments still running are PHENIX and STAR. Two smaller experiments PHOBOS~\cite{PHWP} and BRAHMS~\cite{BRWP} have been completed; a test run, AnDy, occupies the former location of BRAHMS. The LINAC is the injector for polarized protons into the Booster/AGS/RHIC chain; with the Jet Target used for precision beam polarization measurements. The TANDEM injector for Ions has been replaced by the Electron Beam Ion Source (EBIS) starting with the 2012 run. }
\label{fig:RHICoverview}
\end{figure}

\begin{figure}[!h]
\begin{center}
\includegraphics[width=\textwidth]{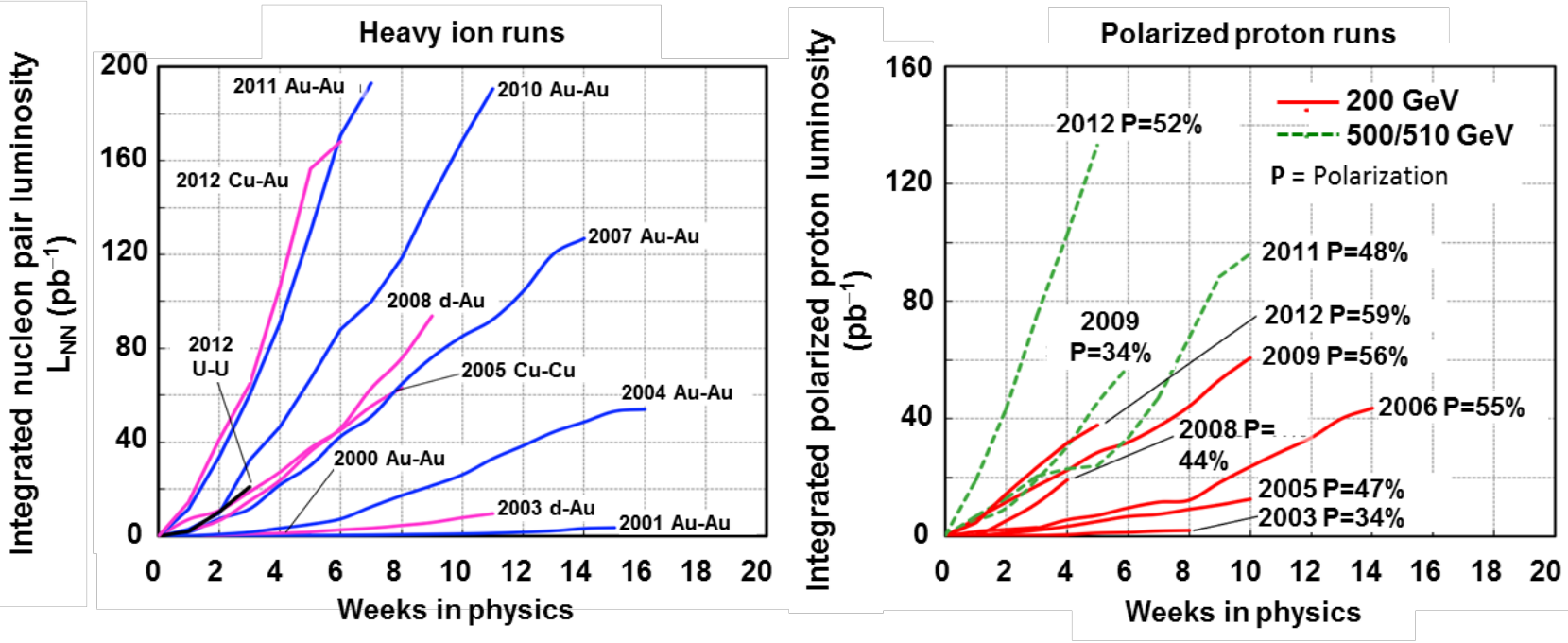}
\end{center}\vspace*{-0.35in}
\caption[]{a)(left) Au+Au performance, where the nucleon-pair luminosity is defined as $L_{\rm NN}=A\times B\times L$, where $L$ is the luminosity and $A$, $B$ are the number of nucleons in the colliding species. b) (right) Polarized p-p performance. Courtesy Wolfram Fischer.}
\label{fig:RHICperf}\vspace*{-2pc}
\end{figure}

With the shutdown of the Tevatron at FERMILAB, on September 30, 2011 after 28 years of operation, RHIC at Brookhaven National Laboratory (Fig.~\ref{fig:RHICoverview}) is the only hadron collider in the U.S. and one of only two hadron-colliders in the world, the other being the CERN-LHC. Also RHIC is the world's first and only polarized proton collider.
RHIC is composed of two independent rings, of circumference 3.8 km, containing a total of 1,740 superconducting magnets. RHIC can collide any species with any other species and since beginning operation in the year 2000 has provided collisions at 15 different values of nucleon-nucleon c.m. energy, $\sqrt{s_{NN}}$, and six different species combinations. 
The performance history of RHIC with A+A and polarized p-p collisions is shown in Fig.~\ref{fig:RHICperf}.  

This past year, the performance was even more outstanding than usual thanks to the new EBIS source and 3-dimensional stochastic cooling~\cite{3DSC-CC1012}. For the first time in a collider, Cu+Au and U+U collisions at $\sqrt{s_{NN}}=200$ GeV were studied, with the purpose of: i) $J/\Psi$ suppression with no-corona in central collisions (Cu+Au); ii) a large eccentricity of the overlap zone in central collisions by using a very large deformed nucleus with a prolate shape, like a rugby ball (U+U).   Also, the polarized p-p runs in 2012 included both $\sqrt{s}=200$ and 510 GeV with  improved polarization and luminosity for the purposes of: iii) comparison data for the new silicon vertex detectors (200 GeV); iv) flavor-identified parton spin distribution functions using the parity violating single spin asymmetry in $W^{\pm}$ production (500 GeV)~\cite{Bunce-ARNPS50}.  

	\section{Issues in A+A versus p-p collisions} 
\begin{figure}[!h]
\begin{center}
\begin{tabular}{cc}
\psfig{file=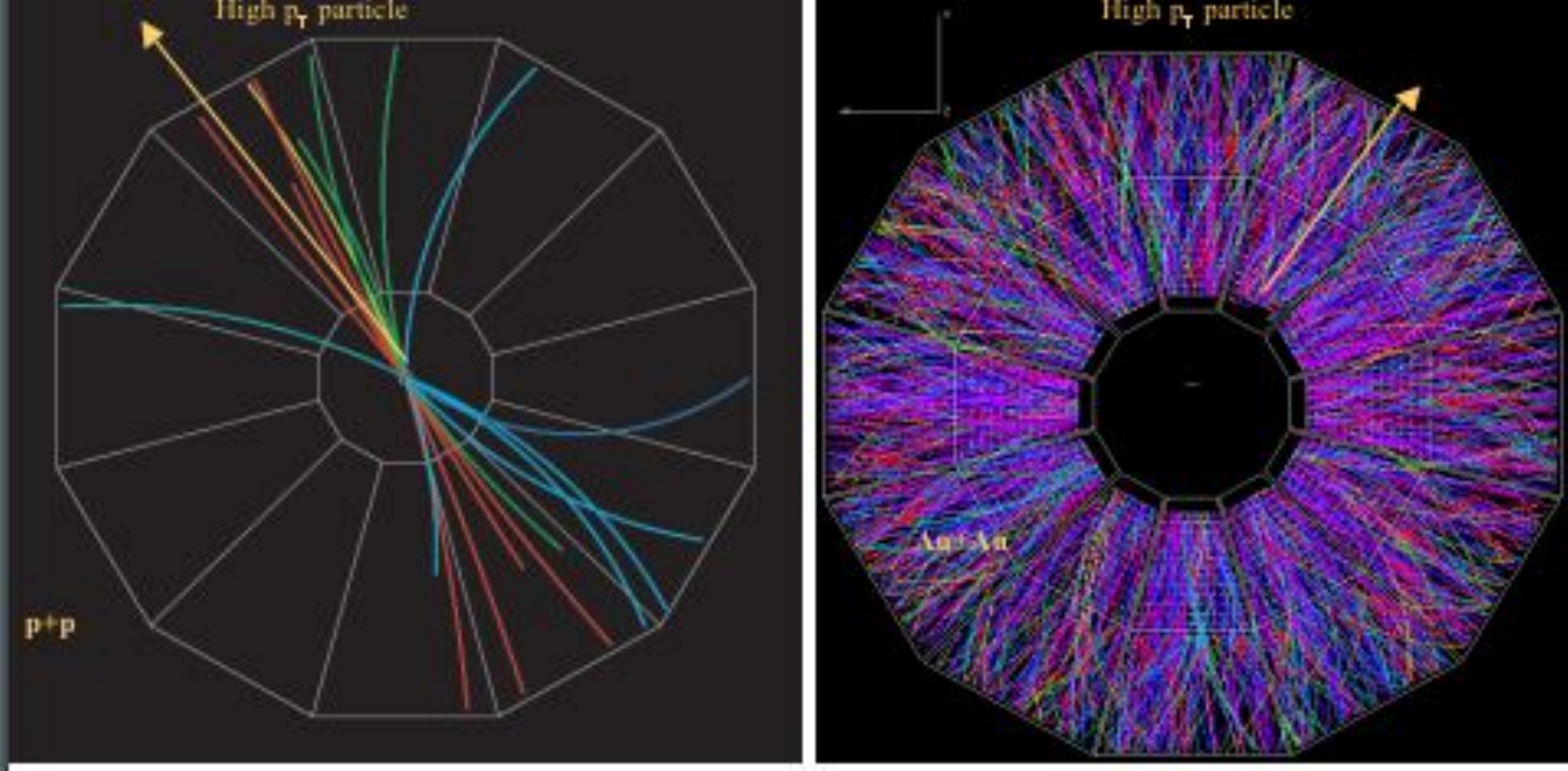,width=0.64\linewidth}&\hspace*{-0.025\linewidth}
\psfig{file=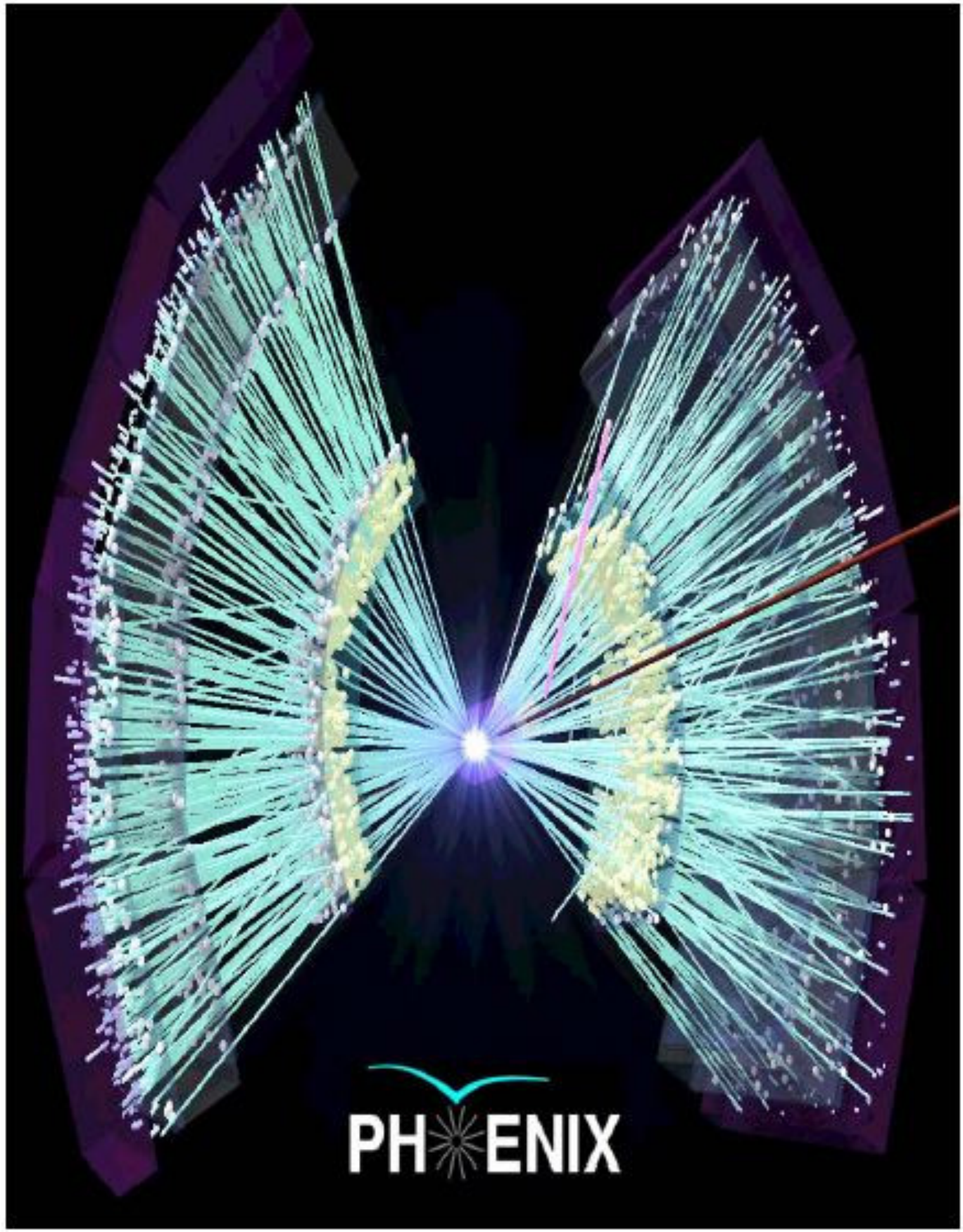,width=0.315\linewidth,height=0.315\linewidth}
\end{tabular}
\end{center}\vspace*{-2pc}
\caption[]{ a) (left) A p-p collision in the STAR detector viewed along the collision axis; b) (center) Au+Au central collision at $\sqrt{s_{NN}}=200$ GeV in STAR;  c) (right) Au+Au central collision at $\sqrt{s_{NN}}=200$ GeV in PHENIX.  
\label{fig:collstar}}\vspace*{-1pc}
\end{figure}

The principal difference in dealing with collisions of relativistic heavy ions, e.g. Au+Au, compared to p-p or e-p (or e-A) collisions at the same nucleon-nucleon c.m. energy, $\sqrt{s_{NN}}$, is that the particle multiplicity is $\sim$ A times larger in A+A central collisions than in p-p collisions as shown in actual events from the STAR and PHENIX detectors at RHIC (Fig.~\ref{fig:collstar}). 
This year, both experiments have benefitted from incremental upgrades, with STAR at present making use of an EM calorimeter with a shower-maximum detector; and PHENIX with central and ``forward'' silicon vertex detectors for event-by-event identification of $c$ and $b$ quarks.  

\begin{figure}[!h]
\begin{center}
\begin{tabular}{cc}
\psfig{file=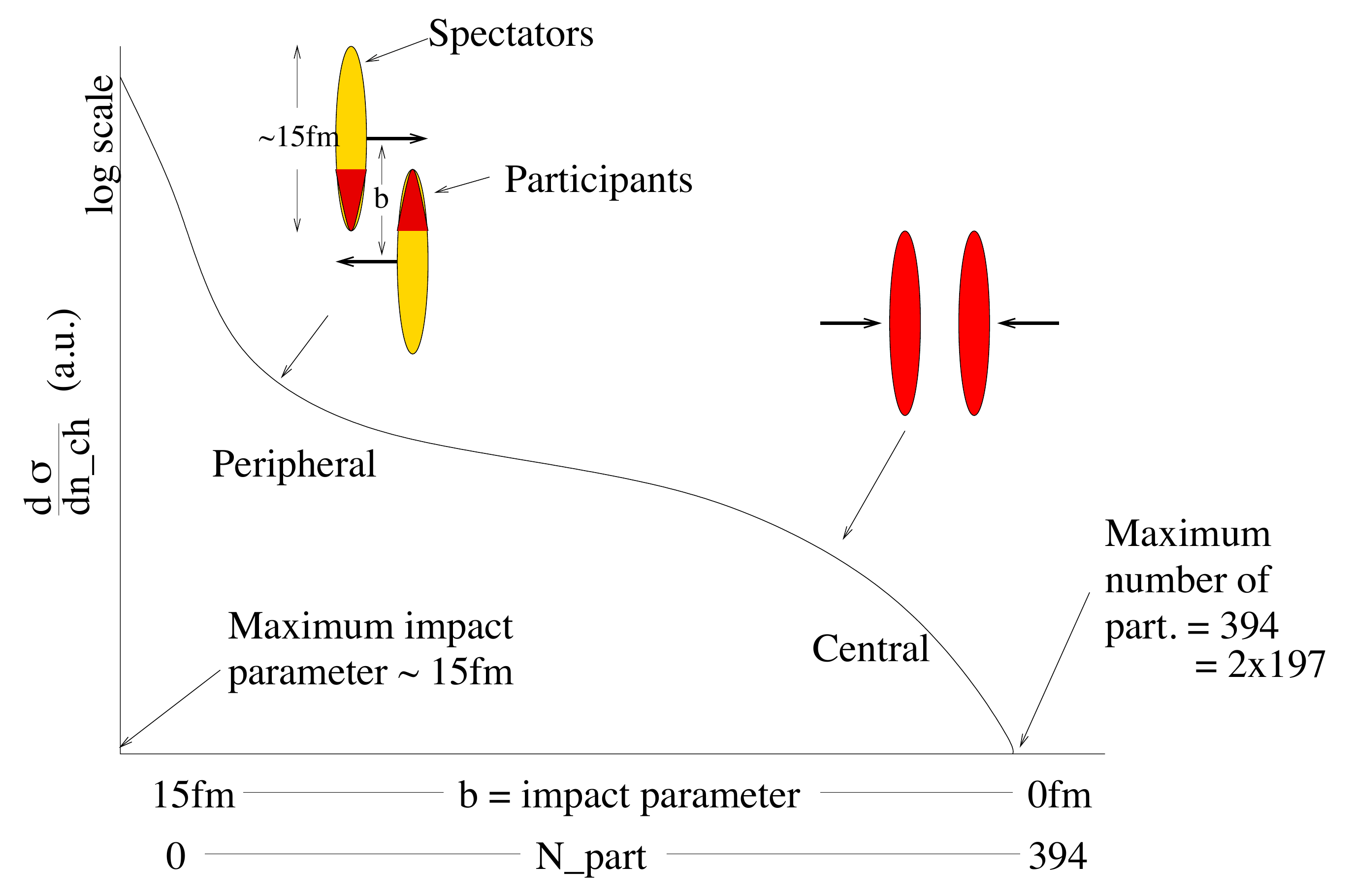,width=0.50\linewidth}\hspace*{-1pc}
\psfig{file=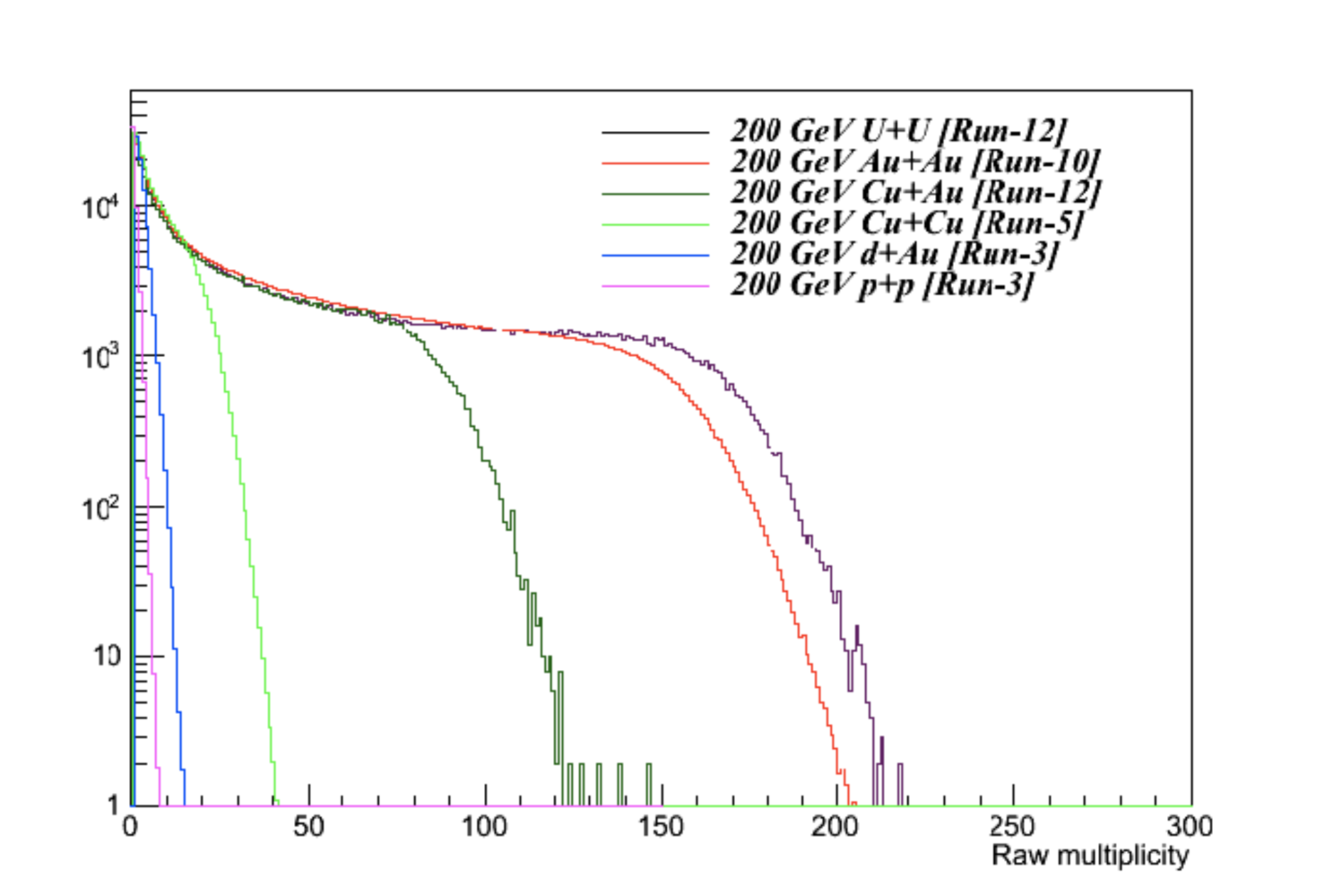,width=0.52\linewidth,angle=0}\end{tabular}
\end{center}\vspace*{-1.5pc}
\caption[]{a) (left) Schematic of collision in the $N$-$N$ c.m. system of two Lorentz contracted nuclei with radius $R$ and impact parameter $b$. The curve with the ordinate labeled $d\sigma/d n_{\rm ch}$ represents the relative probability of charged particle  multiplicity $n_{\rm ch}$ which is directly proportional to the number of participating nucleons, $N_{\rm part}$. b)(right) raw $n_{\rm ch}$ distributions in p-p to U-U collisions at $\sqrt{s_{NN}}=200$ GeV from PHENIX~\cite{JTMQM12}.  
\label{fig:nuclcoll}}\vspace*{-1pc}
\end{figure}

A schematic drawing of a collision of two relativistic Au nuclei is shown in Fig.~\ref{fig:nuclcoll}a. In the center of mass system of the nucleus-nucleus collision, the two Lorentz-contracted nuclei of radius $R$ approach each other with impact parameter $b$. 
In the region of overlap, the ``participating" nucleons interact with each other, while in the non-overlap region, the ``spectator" nucleons simply continue on their original trajectories and can be  measured, in principle, by Zero Degree Calorimeters (ZDC), so that the number of participants, $N_{\rm part}$, can be directly determined. The degree of overlap is called the centrality of the collision, with $b\sim 0$, being the most central and $b\sim 2R$, the most peripheral. The impact parameter $b$ can not be measured directly, so the centrality of a collision is defined in terms of the upper percentile of $n_{\rm ch}$ distributions, e.g. top 10\%-ile, upper $10-20$\%-ile, from which $b$ and $N_{\rm part}$ can be deduced. 

The energy of the inelastic collision is predominantly dissipated by multiple production of soft particles ($\mean{p_T}\approx 0.36$ GeV/c), where $n_{\rm ch}$, the number of charged particles produced, is directly proportional to the number of participating nucleons ($N_{\rm part}$) as sketched on Fig.~\ref{fig:nuclcoll}a. Figure~\ref{fig:nuclcoll}b shows the measured distributions~\cite{JTMQM12} of the charged particle multiplicity, $n_{\rm ch}$, at mid-rapidity at $\sqrt{s_{NN}}=200$ GeV for all the combinations of A+B collisions measured at RHIC. The increase of $n_{\rm ch}$ with both $A$ and $B$ is evident. 

\subsection{Collective Flow}
Another unique feature of A+A collisions compared to either p-p or p+A collisions is the ``flow'' observed, which is a collective effect that can not be obtained from a superposition of independent N-N collisions. 
      \begin{figure}[!thb]
   \begin{center}
\includegraphics[width=0.49\linewidth]{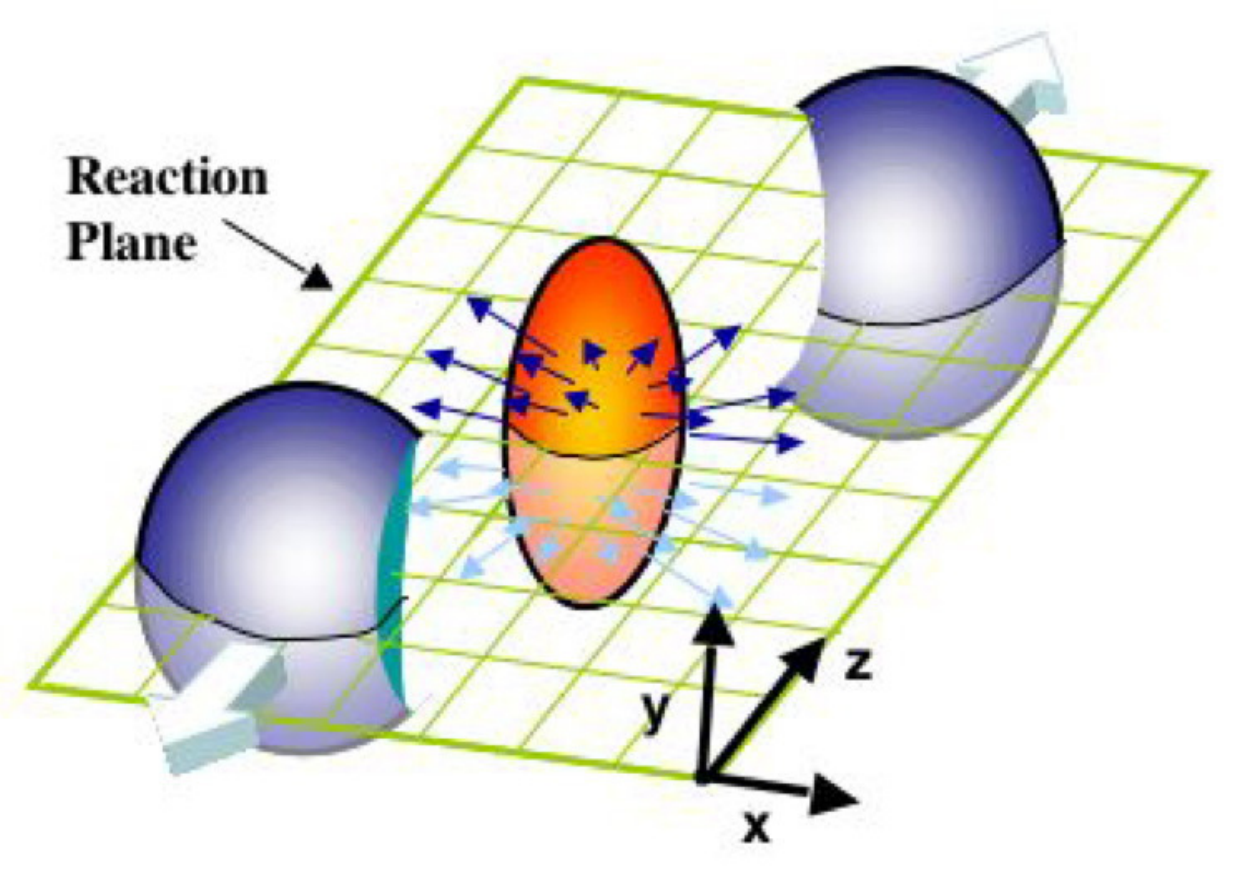}
\includegraphics[width=0.50\linewidth]{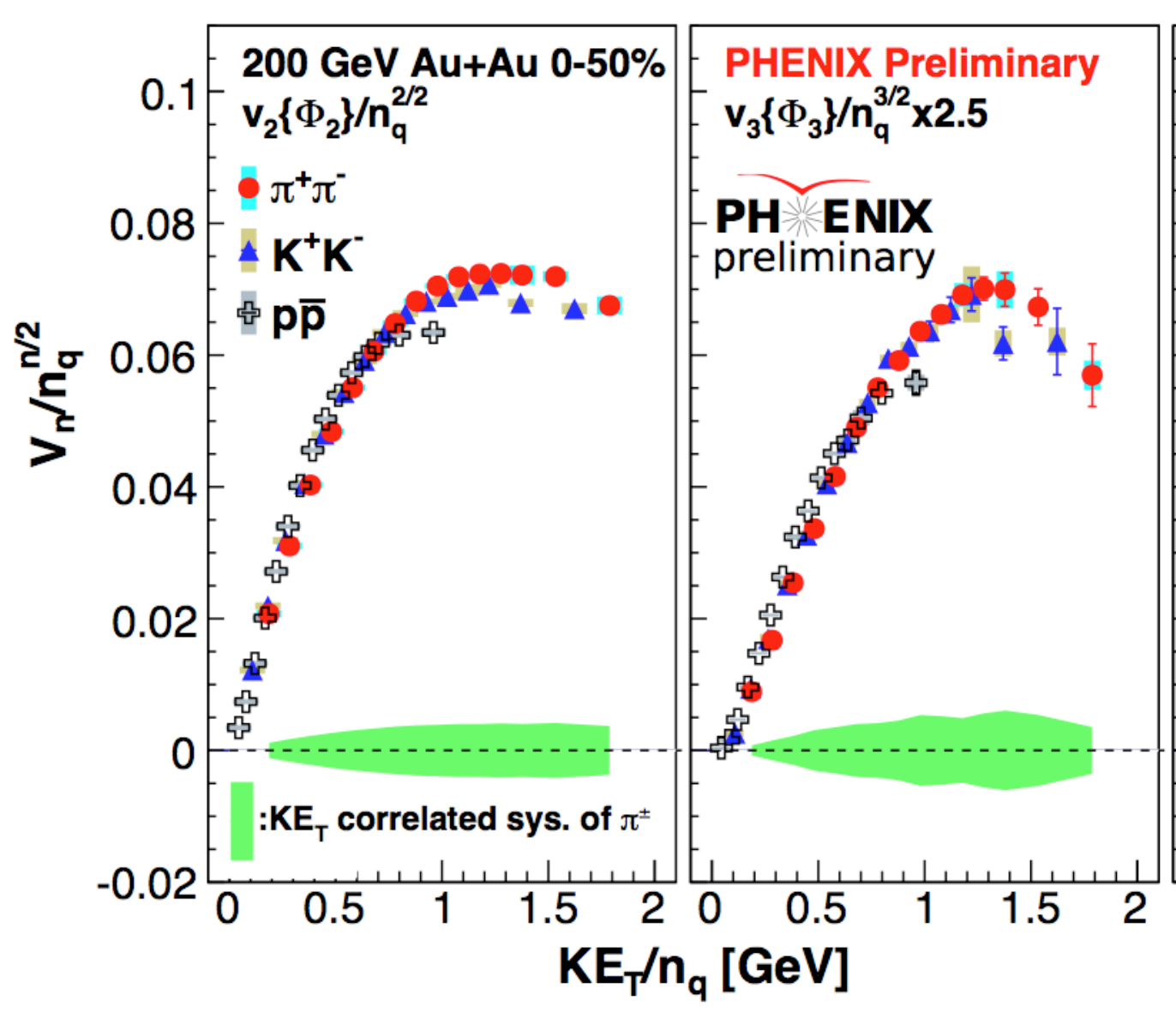}
\end{center}\vspace*{-1pc}
\caption[]{a) (left) Almond shaped overlap zone generated just after an A+A collision where the incident nuclei are moving along the $\pm z$ axis and the impact parameter vector is along the $x$ axis. b)(right) Measurements of elliptical-flow ($v_2$) and triangular flow ($v_3$) for identified hadrons plotted as $v_k$, divided by the number of constituent quarks $n_q$ in the hadron raised to the power ${{k}/{2}}$, as a function of $KE_T/n_q$~\cite{PXYiGuQM12}.   
\label{fig:MasashiFlow}}
\end{figure}
Immediately after an A+A collision, the overlap region defined by the nuclear geometry of participants is roughly almond shaped (see Fig~\ref{fig:MasashiFlow}a); with the shortest axis roughly along the impact parameter vector, which together with the beam ($z$) axis defines the reaction plane. Fluctuations in the distribution of participants from event-to-event do not respect the average symmetry of the almond, $\phi\rightarrow \phi+\pi$, so that the event plane of the participants, with angle $\Psi_k$, is relevant in the fourier decomposition of the azimuthal distribution: 
\begin{equation}
{Ed^3 N}/{dp^3}=({d^3 N}/{2\pi\, p_T dp_T dy}) [1+\sideset{}{_{k\geq1}}\sum 2 v_k \cos k(\phi-\Psi_k)] 
\label{eq:siginv2} 
\end{equation}
which is no longer restricted to even harmonics $v_{k}$ by symmetry so that odd harmonics such as $v_3$ can be produced~\cite{EriceProcPR}, which was observed (Figs~\ref{fig:MasashiFlow}b).   
The flow can also be seen in two-particle correlations.~\footnote{If two particles $A$ and $B$ are correlated to the event  plane but not otherwise correlated, the analog of Eq.~\ref{eq:siginv2} is $dN^{AB}/d\phi^A d\phi^B\propto [1+\sum_{k\geq1} 2 v^A_k v^B_k\cos k(\phi^A-\phi^B)]$.}

The fact that the flow persists for $p_T>1$ GeV/c~\cite{TeaneyPRC68} and that the higher harmonics are not strongly damped~\cite{AlverOllitrault,Mishra08} implies that the viscosity is small, perhaps as small as a quantum viscosity bound from string theory~\cite{Kovtun05}, $\eta/s=1/(4\pi)$ where $\eta$ is the shear viscosity and $s$ the entropy density per unit volume.  This has led to the description of the ``s\QGP'' produced at RHIC as ``the perfect fluid''~\cite{THWPS}. 
It was observed that the flow is not dominated by final state hadrons but is proportional to the number of constituent quarks $n_q$ in the mesons and baryons, in which case $v_2/n_q$, as well as $v_3/n_q^{3/2}$~\cite{PXYiGuQM12}, as a function of the transverse kinetic energy per constituent quark, $KE_T/n_q$, would be universal, as shown in Fig.~\ref{fig:MasashiFlow}b,  where the power $k/2$ and the transverse kinetic energy, $KE_T\equiv m_T-m_0$, are suggested by the relativistic hydrodynamics of ``ideal'' fluids~\cite{BorghiniOllitrault,LaceyQM11}. 

\section{Old and New RHIC results}
   One of the most important innovations at RHIC was the use of hard scattering as an in-situ probe of the medium produced in A+A collisions by the effect of the medium on outgoing hard-scattered partons produced by the    \centerline{\mbox{\includegraphics[width=0.9\textwidth]{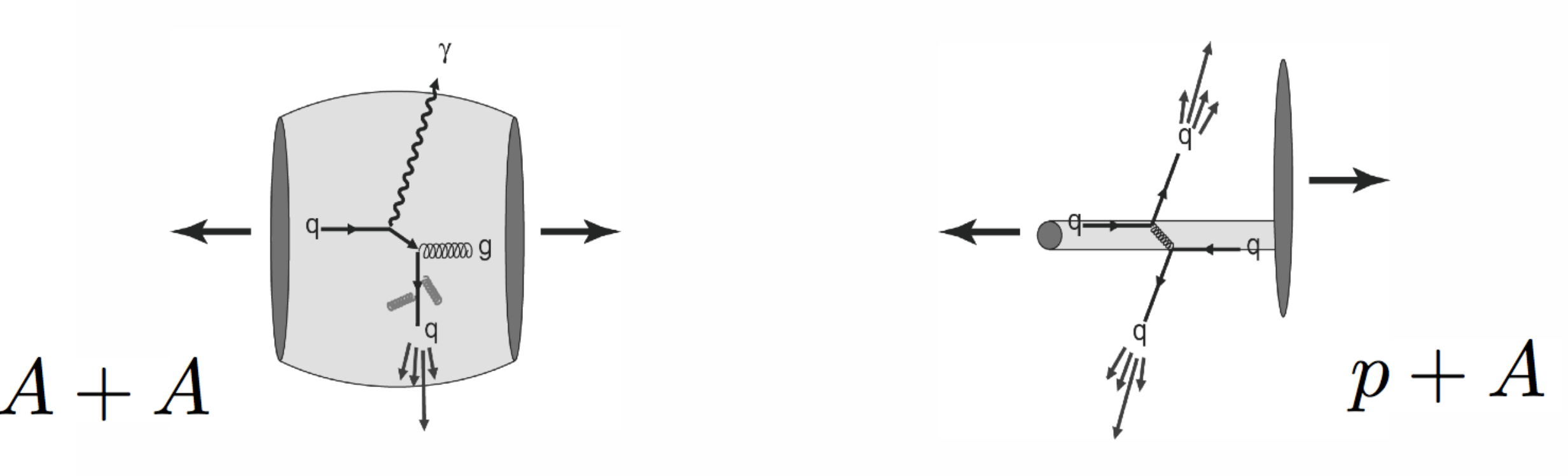} }} 
\mbox{initial}  A+A collision. Measurements in p+A (or d+A) collisions, where no (or negligible) medium is produced, allow correction for any modification of the nuclear structure function from an incoherent superposition of proton and neutron structure functions. 

\begin{figure}[hbt]
\begin{center}
\includegraphics[width=0.34\textwidth]{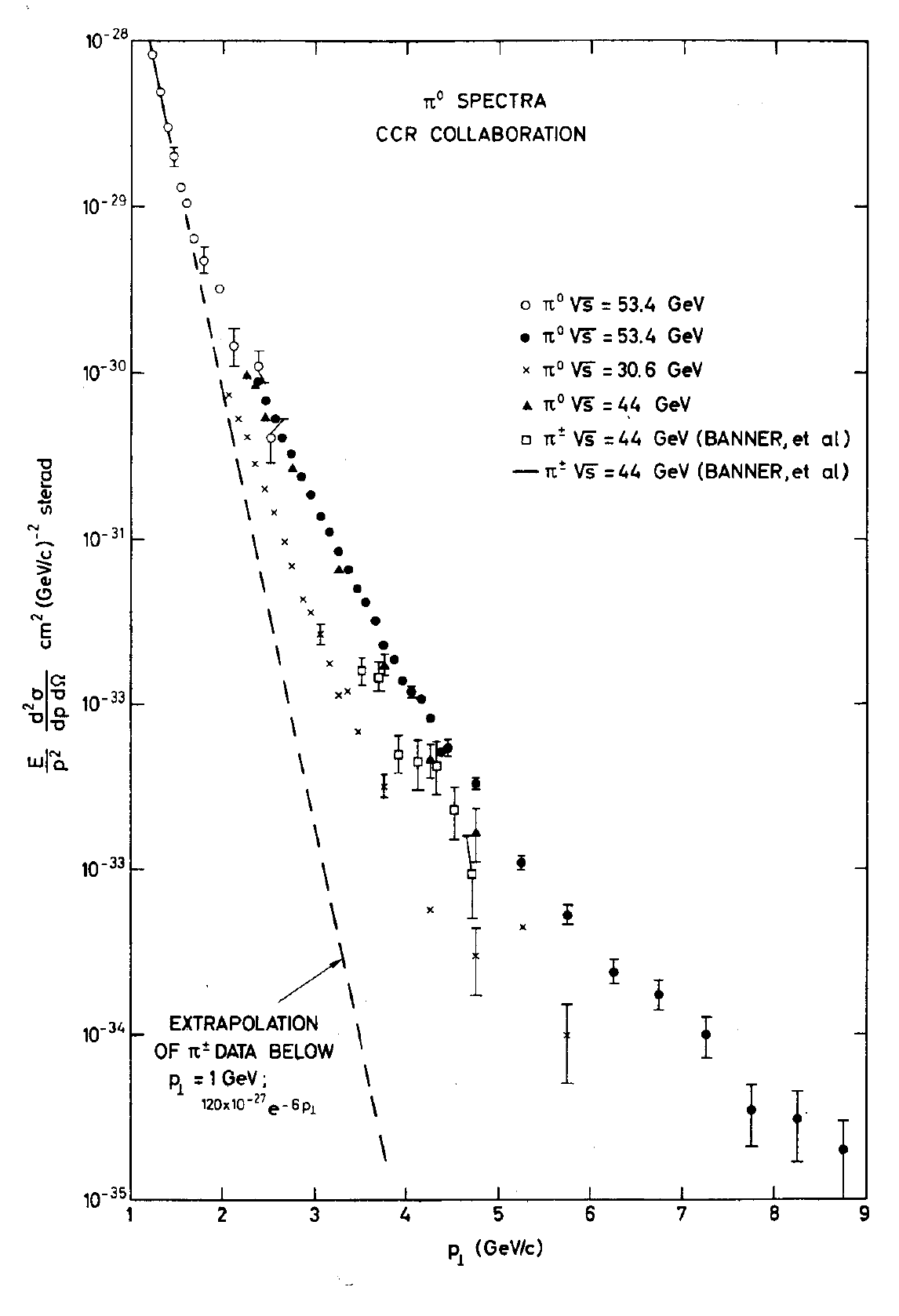}
\includegraphics[width=0.32\textwidth,height=0.48\textwidth]{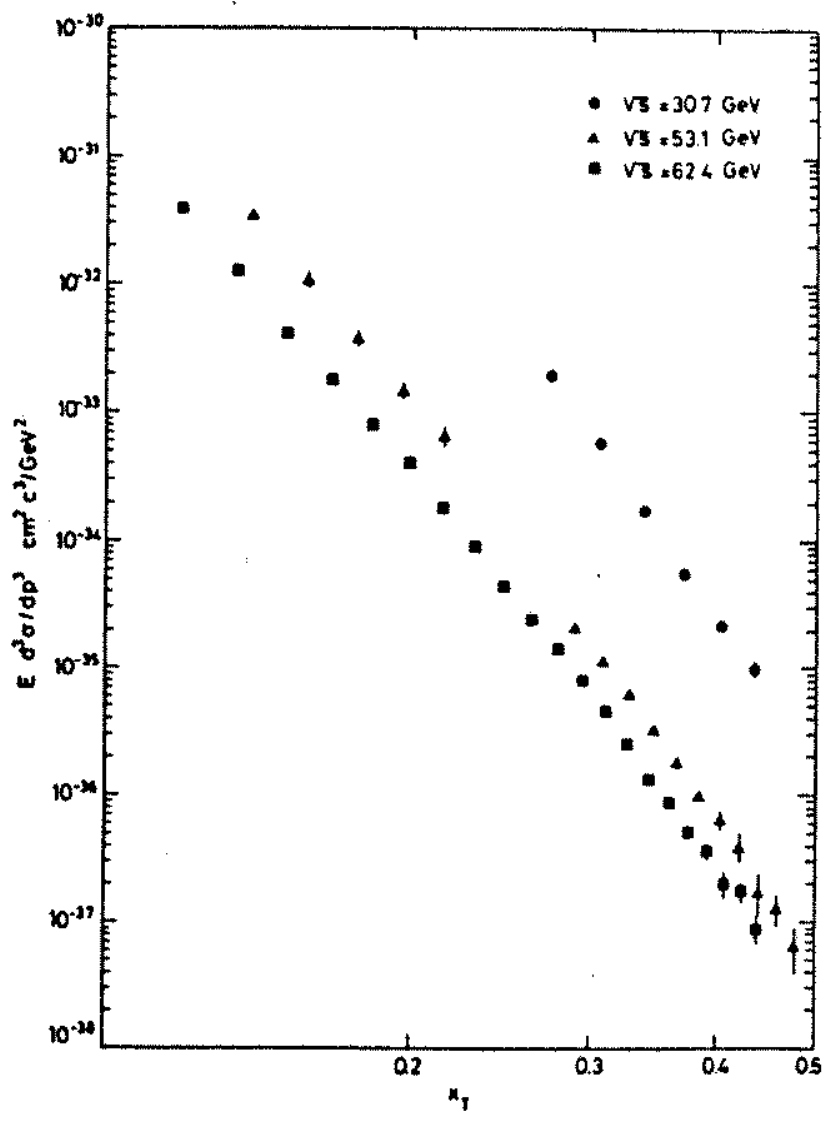}
\includegraphics[width=0.32\textwidth,height=0.32\textwidth]{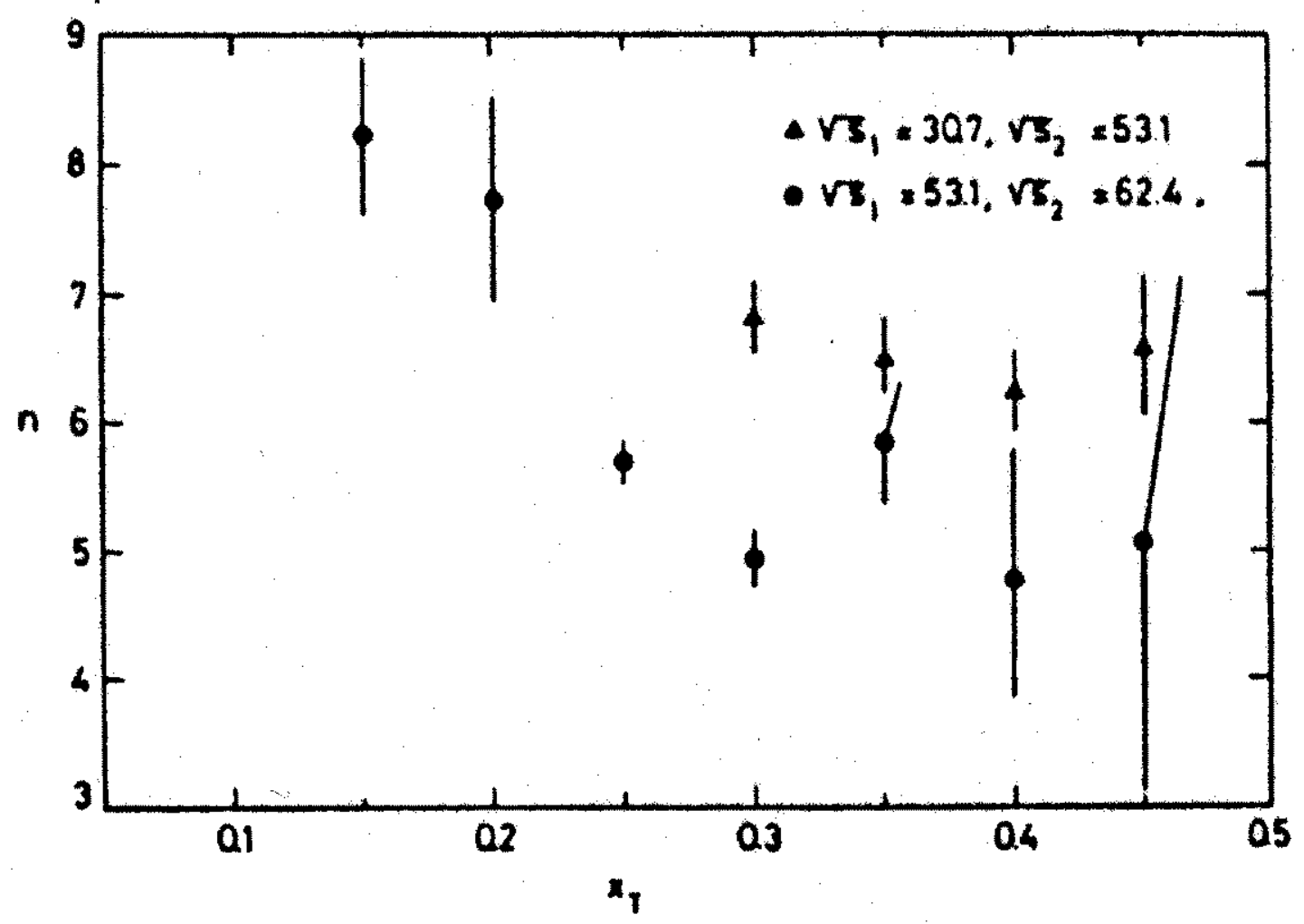}

\end{center}
\vspace*{-0.28in}
\caption[]{a) (left) Plot of invariant single particle inclusive cross sections vs. $p_T$ at the CERN-ISR~\cite{CoolICHEP72}. b) (center) Log-log plot of CCOR~\cite{CCOR} invariant cross sections vs $x_T=2 p_T/\sqrt{s}$. c) (right) CCOR~\cite{CCOR} $n_{\rm eff}(x_T,\sqrt{s})$ derived from the combinations indicated.  The systematic normalization error at $\sqrt{s}=30.6$ GeV has been added in quadrature. There is an additional common systematic error of $\pm 0.33$ in $n_{\rm eff}$.  
\label{fig:ccorxt} }
\end{figure}

Hard-scattering of partons in p-p collisions was discovered at the CERN-ISR in 1972 as shown at the ICHEP 1972 conference~\cite{CoolICHEP72} by observation that the $e^{-6p_T}$ dependence for pion production at low $p_T$ breaks to a power-law for $p_T>2$ GeV/c with a characteristic $\sqrt{s}$ dependence (Fig.~\ref{fig:ccorxt}a) which is more evident from the log-log plot of subsequent $\pi^0$ data (Fig~\ref{fig:ccorxt}b). This plot, as a function of $x_T=2p_T/\sqrt{s}$ (Fig~\ref{fig:ccorxt}b), shows that the cross section for hard-processes (Fig~\ref{fig:ccorxt}c) obeys the scaling law, called``$x_T$-scaling'':
\begin{equation}
E{{ d^3\sigma} \over {d^3p}}={1 \over p_T^{\,n_{\rm eff}} } F ({p_T \over \sqrt{s} })={1 \over \sqrt{s}^{\,n_{\rm eff}} } G({x_T} ) 
\end{equation}
for $\sqrt{s}\geq 53.1$ GeV, $x_T\geq 0.3$ ($p_T\geq 7$ GeV/c), where $n_{\rm eff}(x_T,\sqrt{s})\sim 4-6$ gives the form of the force-law between constituents as predicted by Quantum Chromodynamics (QCD) with non-scaling structure and fragmentation functions and running coupling constant~\cite{EriceProcPR}.  

Figure~\ref{fig:ccorxt}a provided the first evidence that the electrically charged partons of e-p deeply inelastic scattering (DIS) interacted with each other much more strongly than electromagnetically~\cite{JanMikeBook}. Note that the parton degrees of freedom in p-p collisions become evident only for parton-parton scattering at large $Q^2\approx 2p_T^2\,\gsim\, 5-8$ (GeV/c)$^2$, while in DIS, only $Q^2>1$ (GeV/c)$^2$ is required. 

The use of hard-scattering to probe the thermal or ``soft'' medium produced in RHI collisions was stimulated by pQCD studies~\cite{BDMPS} of the energy loss of partons produced by hard scattering, ``with their color charge fully exposed'', in traversing a medium ``with a large density of similarly exposed color charges''. The conclusion was that ``Numerical estimates of the loss suggest that it may be significantly greater in hot matter than in cold. {\em This makes the magnitude of the radiative energy loss a remarkable signal for \QGP\  formation}''~\cite{BDMPS}. In addition to being a probe of the \QGP\ the fully exposed color charges allow the study of parton-scattering with $Q^2 \ll 1-5$ (GeV/c)$^2$ in the medium where new collective \QCD\ effects may possibly be observed.

\begin{figure}[!t]
\includegraphics[height=0.26\textheight]{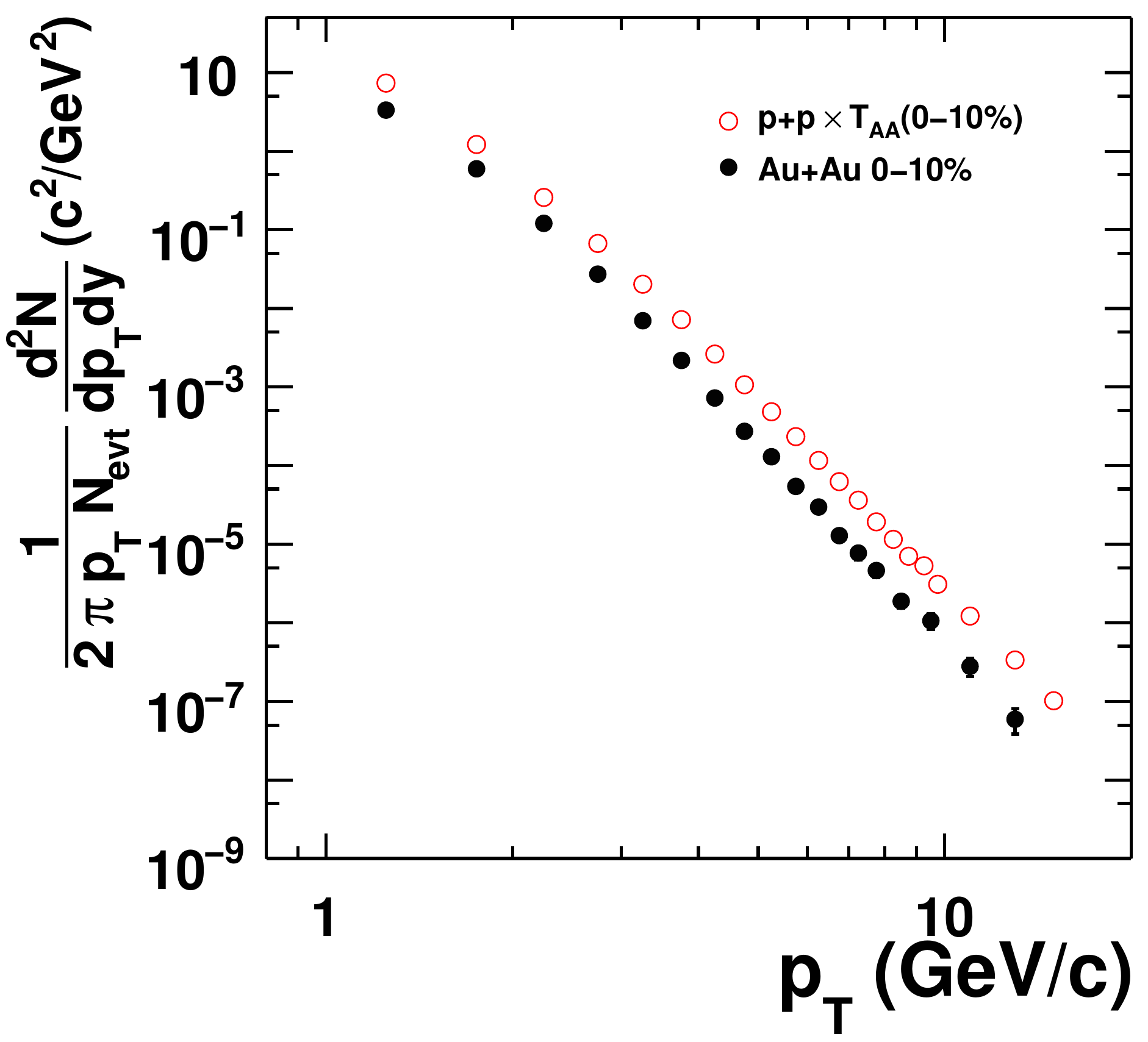}
\hspace*{0.001\textwidth} \includegraphics[height=0.26\textheight]{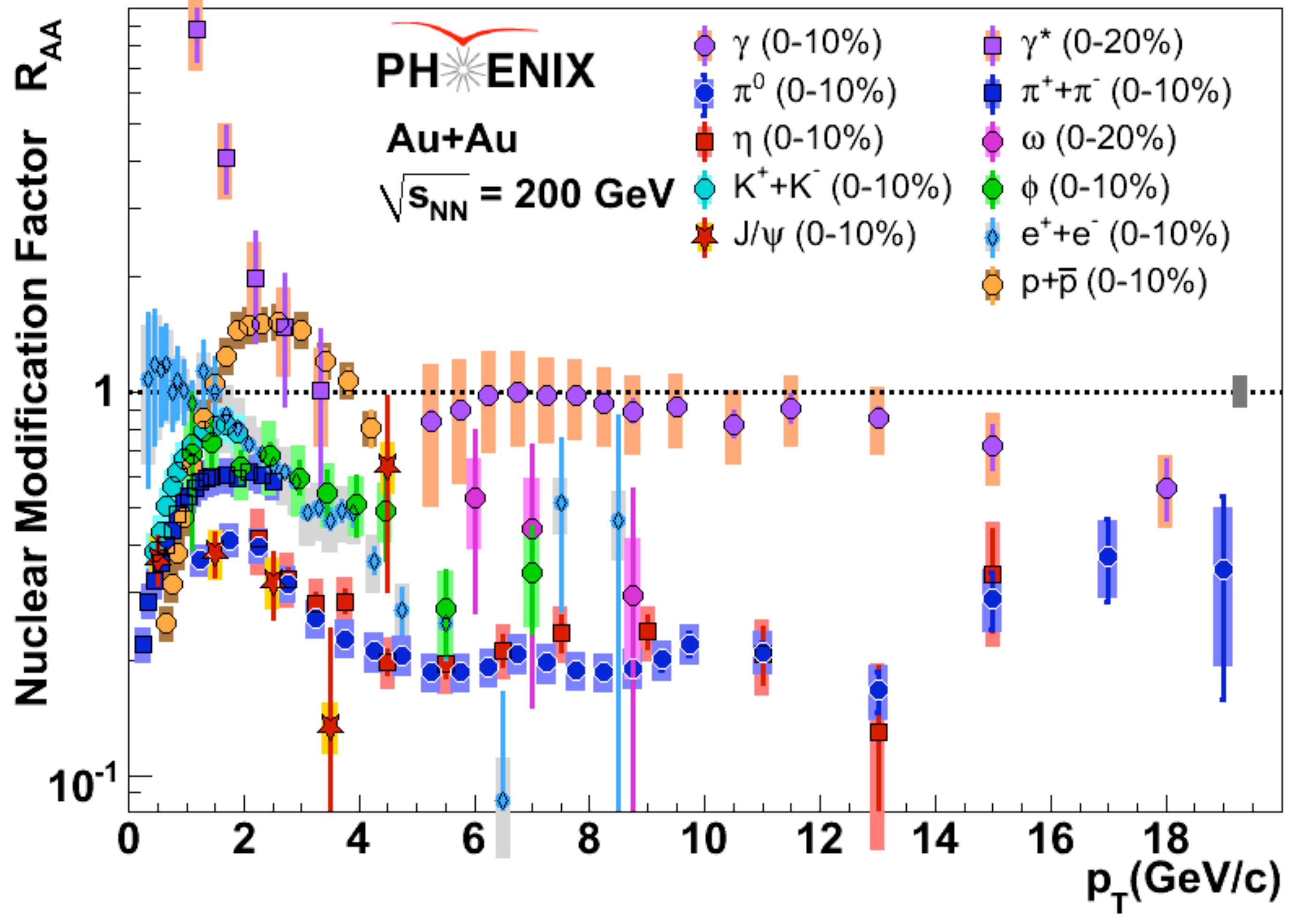}\vspace*{-0.5pc}
\caption{a) (left) Log-log plot of invariant yield of $\pi^0$ at $\sqrt{s_{NN}}=200$ GeV as a function of transverse momentum $p_T$ in p-p collisions multiplied by $\mean{T_{AA}}$ for Au+Au central (0--10\%) collisions compared to the Au+Au measurement~\cite{ppg054}. b) (right) $R_{AA}(p_T)$ for all identified particles so far measured by PHENIX in Au+Au central collisions at $\sqrt{s_{NN}}=200$ GeV.}
\label{fig:Tshirt}\vspace*{-0.5pc}
\end{figure}

   The discovery, at RHIC~\cite{ppg003}, that $\pi^0$'s produced at large transverse momenta are suppressed in central Au+Au collisions by a factor of $\sim5$ compared to point-like scaling from p-p collisions is arguably {\em the}  major discovery in Relativistic Heavy Ion Physics. For $\pi^0$ (Fig.~\ref{fig:Tshirt}a)~\cite{ppg054} the hard-scattering in p-p collisions is indicated by the power law behavior $p_T^{-n}$ for the invariant cross section, $E d^3\sigma/dp^3$, with $n=8.1\pm 0.05$ for $p_T\geq 3$ GeV/c.  The Au+Au data at a given $p_T$ can be characterized either as shifted lower in $p_T$ by $\delta p_T$ from the point-like scaled p-p data at $p'_T=p_T+\delta p_T$, or shifted down in magnitude, i.e. suppressed. In Fig.~\ref{fig:Tshirt}b, the suppression of the many identified particles measured by PHENIX at RHIC is presented as the Nuclear Modification Factor, 
$R_{AA}(p_T)$, the ratio of the yield of e.g. $\pi$ per central Au+Au collision (upper 10\%-ile of observed multiplicity)  to the point-like-scaled p-p cross section, where $\mean{T_{AA}}$ is the average overlap integral of the nuclear thickness functions: 
   \begin{equation}
  R_{AA}(p_T)=\frac{{d^2N^{\pi}_{AA}/dp_T dy N_{AA}}} { \mean{T_{AA}} d^2\sigma^{\pi}_{pp}/dp_T dy} \quad . 
  \label{eq:RAA}
  \end{equation}

The striking differences of $R_{AA}(p_T)$ in central Au+Au collisions for the many particles measured by PHENIX  (Fig.~\ref{fig:Tshirt}b) illustrates the importance of particle identification for understanding the physics of the medium produced at RHIC. Most notable are: the equal suppression of $\pi^0$ and $\eta$ mesons by a constant factor of 5 ($R_{AA}=0.2$) for $4\leq p_T \leq 15$ GeV/c, with suggestion of an increase in $R_{AA}$ for $p_T > 15$ GeV/c; the equality of suppression of direct-single $e^{\pm}$ (from heavy quark ($c$, $b$) decay) and $\pi^0$ at $p_T\gsim 5$ GeV/c; the non-suppression of direct-$\gamma$ for $p_T\geq 4$ GeV/c; the exponential rise of $R_{AA}$ of direct-$\gamma$ for $p_T<2$ GeV/c~\cite{ppg086}, which is totally and dramatically different from all other particles and attributed to thermal photon production by many authors (e.g. see citations in reference~\cite{ppg086}). For $p_T\gsim 4$ GeV/c, the hard-scattering region,  the fact that all hadrons are suppressed, but direct-$\gamma$ are not suppressed, indicates that suppression is a medium effect on outgoing color-charged partons likely due to energy loss by coherent Landau-Pomeranchuk-Migdal radiation of gluons, predicted in pQCD~\cite{BDMPS}, which is sensitive to properties of the medium. Measurements of two-particle correlations~\cite{EriceProcPR} confirm the loss of energy of the away-jet relative to the trigger jet in Au+Au central collisions compared to p-p collisions. However, there are still many details which remain to be understood, such as the apparent suppression of direct-$\gamma$ for $p_T\gsim\ 18$ GeV/c, approaching that of the $\pi^0$. Interesting new results this year extend and clarify these observations. 
   
\begin{figure}[!b]
   \begin{center}
\includegraphics[width=0.49\textwidth,height=0.38\textwidth]{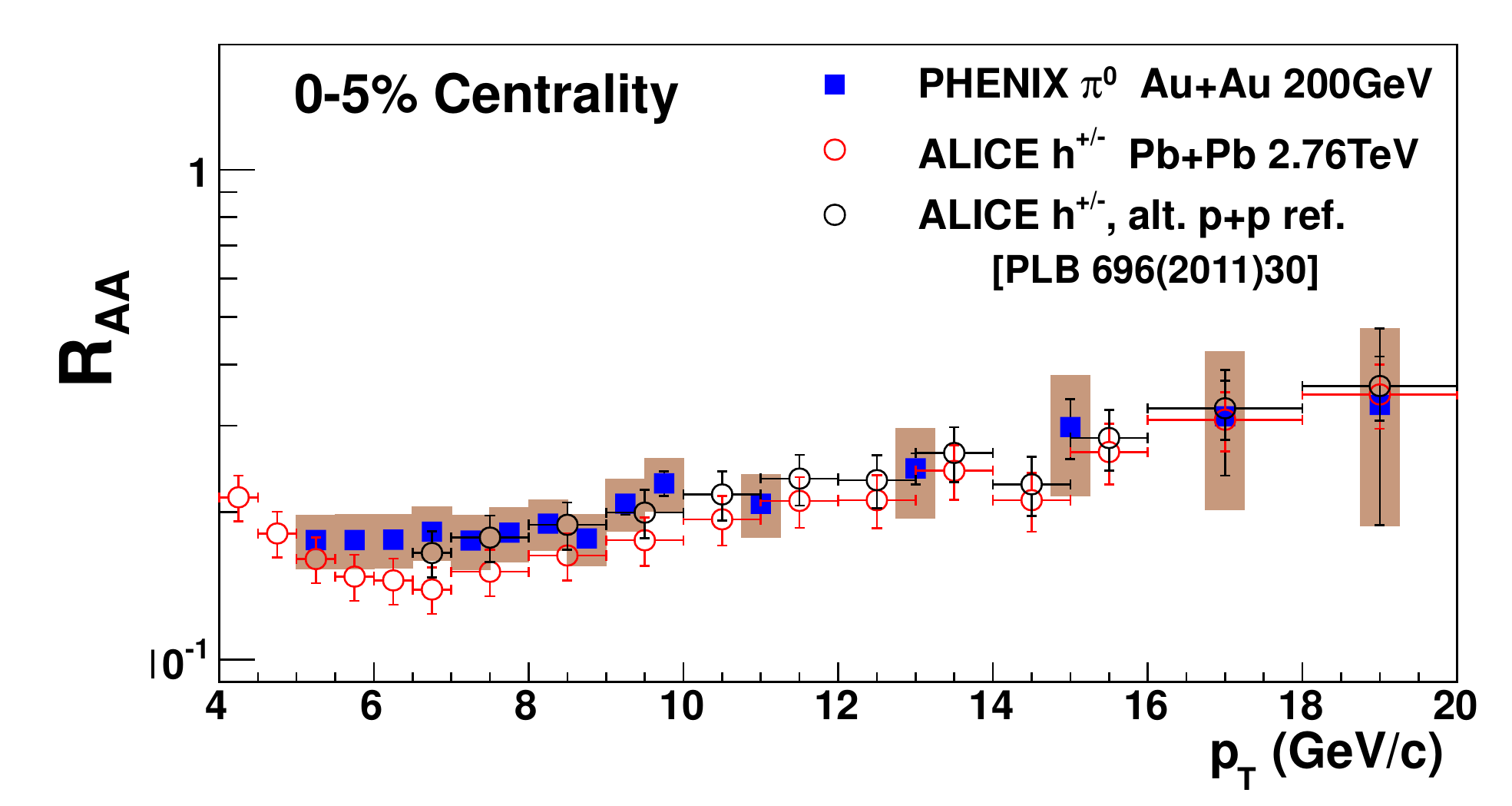}
\includegraphics[width=0.49\textwidth]{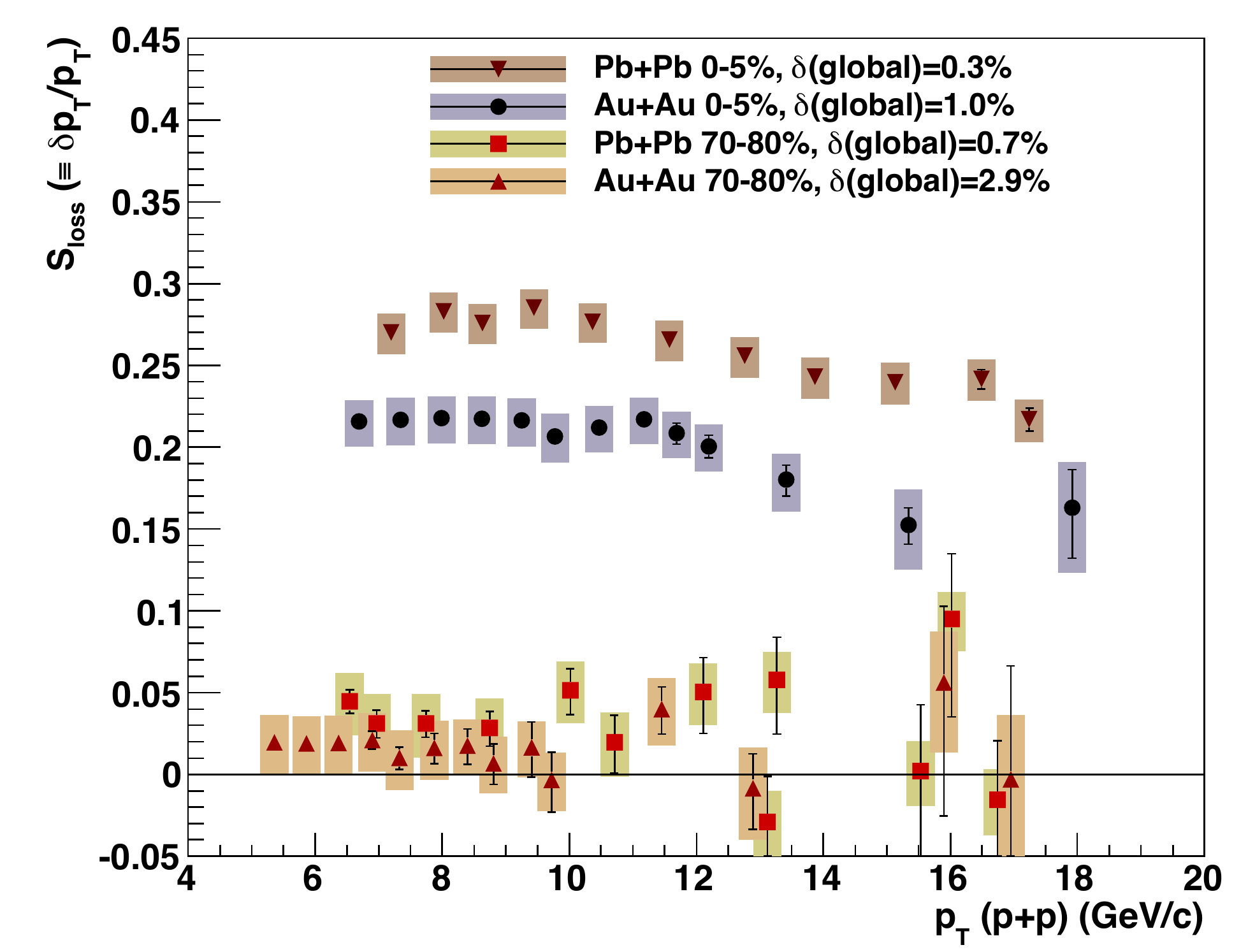}
\end{center}\vspace*{-0.25in}
\caption[]{a) (left) $R_{AA}$ of $\pi^0$ in $\sqrt{s_{NN}}=200$ GeV central (0-5\%) Au+Au collisions~\cite{ppg133} at RHIC compared to charged hadron $R_{AA}$ in $\sqrt{s_{NN}}=2.76$ TeV central (0-5\%)Pb+Pb collisions at LHC. b) (right) Shift of $p_T$ spectrum $S_{\rm loss}=\delta p_T/p_T$ vs. $p_T$ (p-p) calculated by PHENIX~\cite{ppg133} for RHIC and LHC. 
\label{fig:ppg133}}\vspace*{-1pc}
\end{figure}

An improved measurement of $\pi^0$ production in Au+Au and p-p collisions by PHENIX~\cite{ppg133} now clearly shows a significant increase of $R_{AA}$ (decrease in suppression) with increasing $p_T$ over the range $7<p_T<20$ GeV/c  for 0-5\% central Au+Au collisions at $\sqrt{s_{NN}}=200$ GeV (Fig.~\ref{fig:ppg133}a).  
Interestingly, despite more than a factor of 20 higher $\sqrt{s_{NN}}$, the ALICE $R_{AA}$ data from LHC~\cite{ALICEPLB696} are nearly identical to the RHIC measurement for $5< p_T <20$ GeV/c. However, since the exponent of the power-law at LHC ($n\sim 6$) is flatter than that at RHIC ($n\approx 8$, Fig.~\ref{fig:Tshirt}a), a $\sim 25$\% larger  shift $\delta p_T/p_T$ in the spectrum from p-p to A+A is required at LHC to get the same $R_{AA}$ (Fig.~\ref{fig:ppg133}b), likely indicating $\sim 25$\% larger fractional energy loss at LHC in this $p_T$ range.   
 
\begin{figure}[!h]
   \begin{center}
\includegraphics[width=0.49\textwidth]{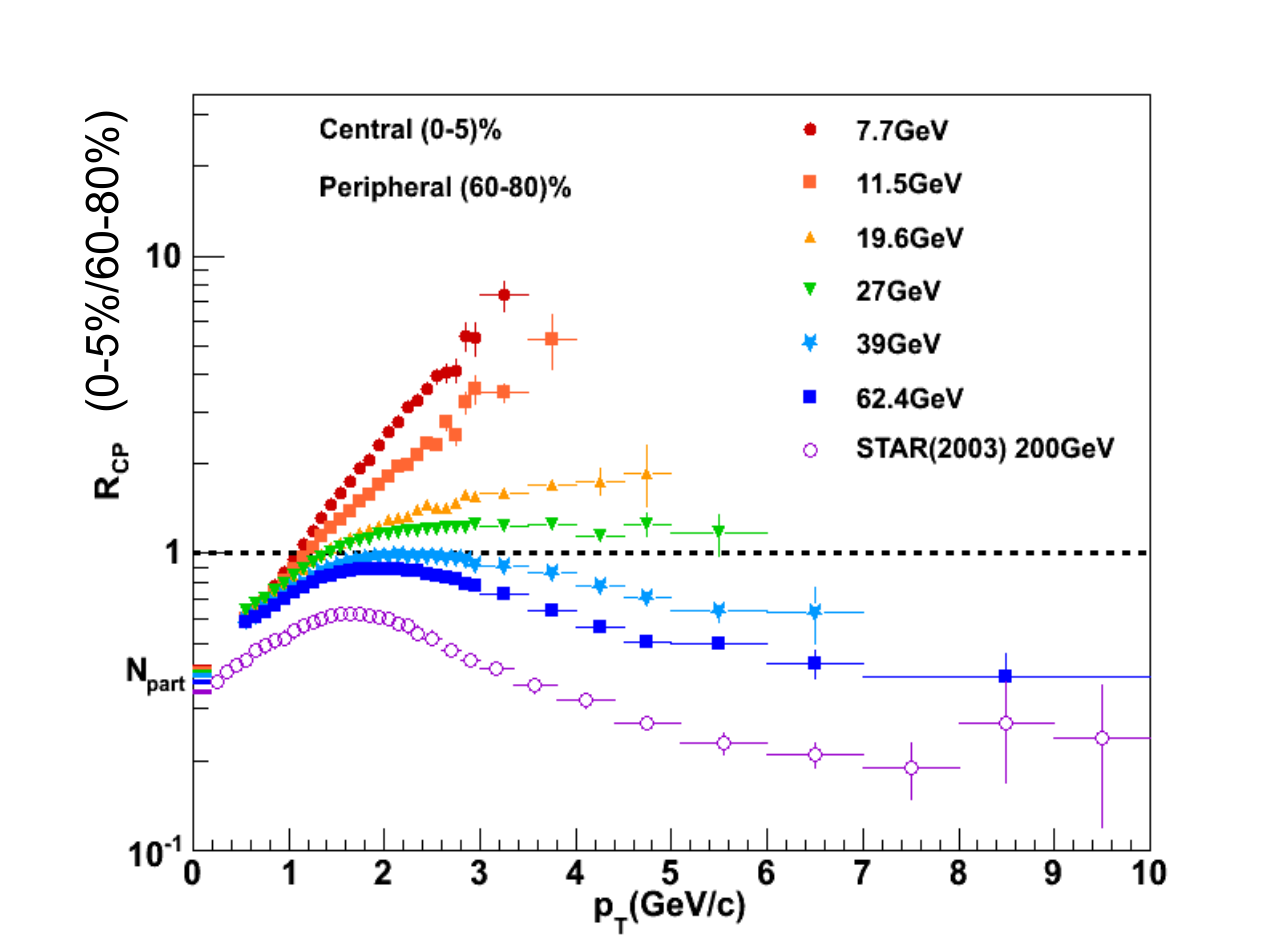}
\includegraphics[width=0.49\textwidth,height=0.30\textwidth]{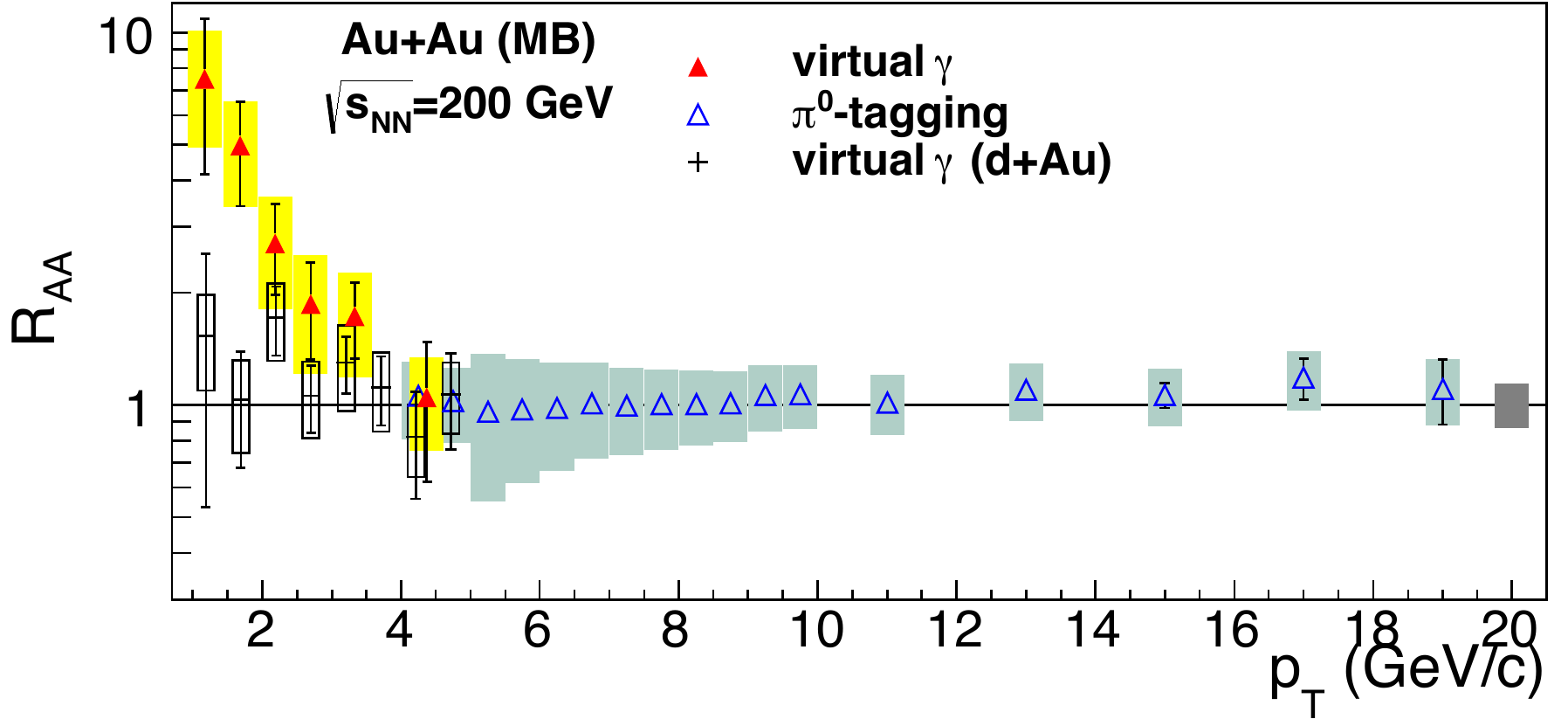}
\end{center}\vspace*{-0.25in}
\caption[]{a) (left) STAR $R_{\rm CP} (p_T)$ for $h^{\pm}$ as a function of $\sqrt{s_{NN}}$ in Au+Au collisions~\cite{STARhiQM2012}. b)(right) PHENIX $R_{AA} (p_T)$ of direct-$\gamma$ in d+Au and Au+Au minimum bias collisions at $\sqrt{s_{NN}}=200$ GeV~\cite{ppg140}.   
\label{fig:twoRAA}}
\end{figure}

Measurements by STAR (Fig.~\ref{fig:twoRAA}a)~\cite{STARhiQM2012} of the evolution of charged hadron suppression with $\sqrt{s_{NN}}$ in Au+Au collisions, using the variable $R_{CP}=(R_{AA}^{0-5\%}/R_{AA}^{60-80\%})$ which does not require the p-p cross section (see Eq.~\ref{eq:RAA}) but is usually smaller than $R_{AA}$~\cite{BRWP}, show the transition from suppression ($R_{\rm CP}<1$) to enhancement ($R_{\rm CP}>1$) for $\sqrt{s_{NN}} > 27$ GeV.          

Improved measurements of direct-$\gamma$ production in p-p, dAu and Au+Au collisions by PHENIX~\cite{ppg140} show several interesting results. In Fig.~\ref{fig:twoRAA}b, new measurements of $R_{AA}\approx 1$ for d+Au using internal conversions in the thermal region, $p_T<4$ GeV,  reinforce the uniqueness of the exponential rise of the Au+Au minimum bias photon spectrum, thus confirming that the exponential for $p_T<4$ GeV/c in Au+Au is a hot matter effect, i.e. thermal photon production.   Also in Fig.~\ref{fig:twoRAA}b, improved measurements of real direct-$\gamma$ in Au+Au collisions, by eliminating background from $\gamma$-rays associated with a second $\gamma$ in the $\pi^0$ mass range ($\pi^0$ tagging), no longer show an ``apparent suppression'' but are consistent with $R_{AA}=1$ out to 20 GeV/c.

\begin{figure}[!t]
   \begin{center}
\includegraphics[width=0.49\textwidth]{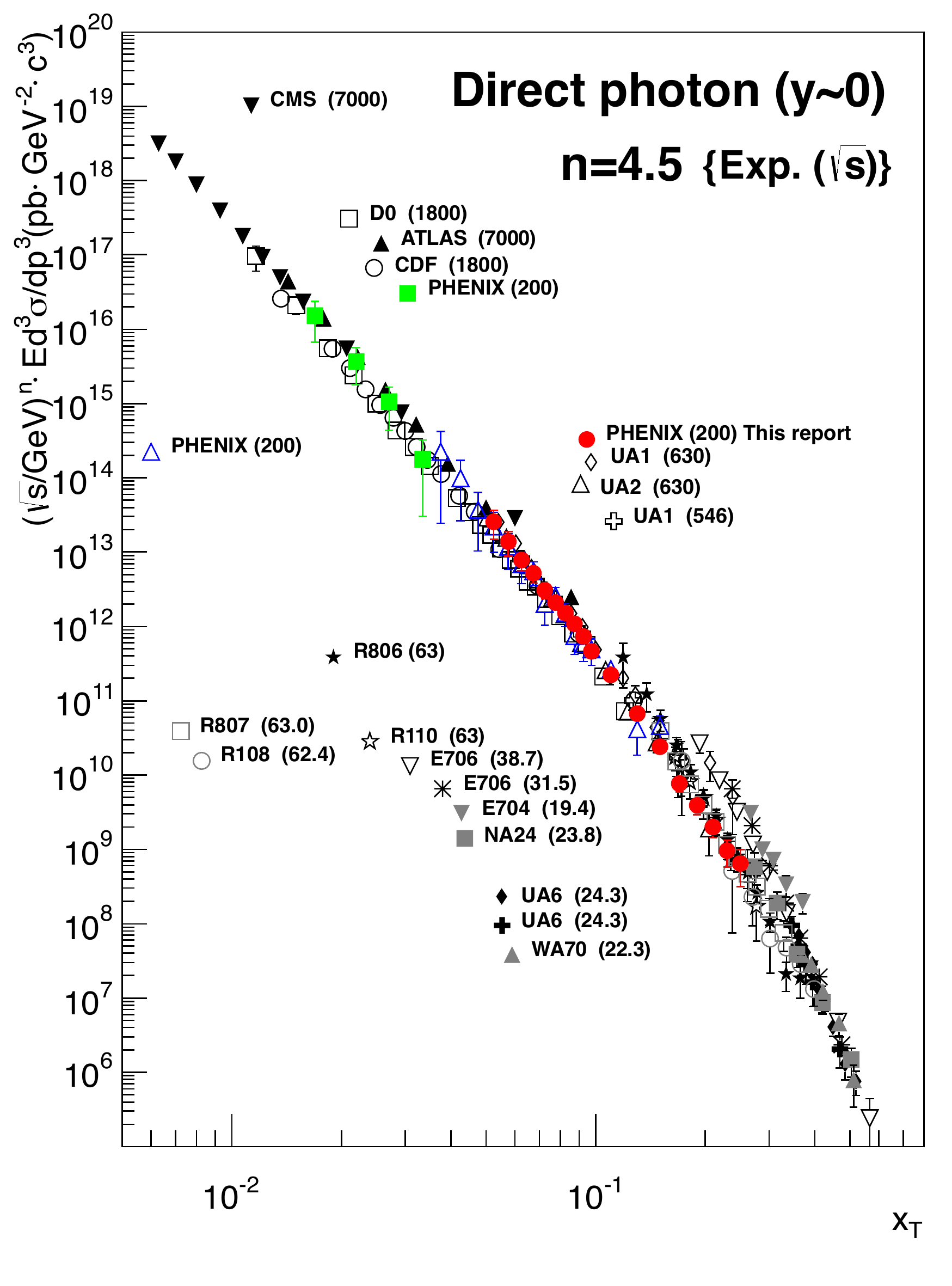}
\raisebox{0.2pc}{\includegraphics[width=0.492\textwidth]{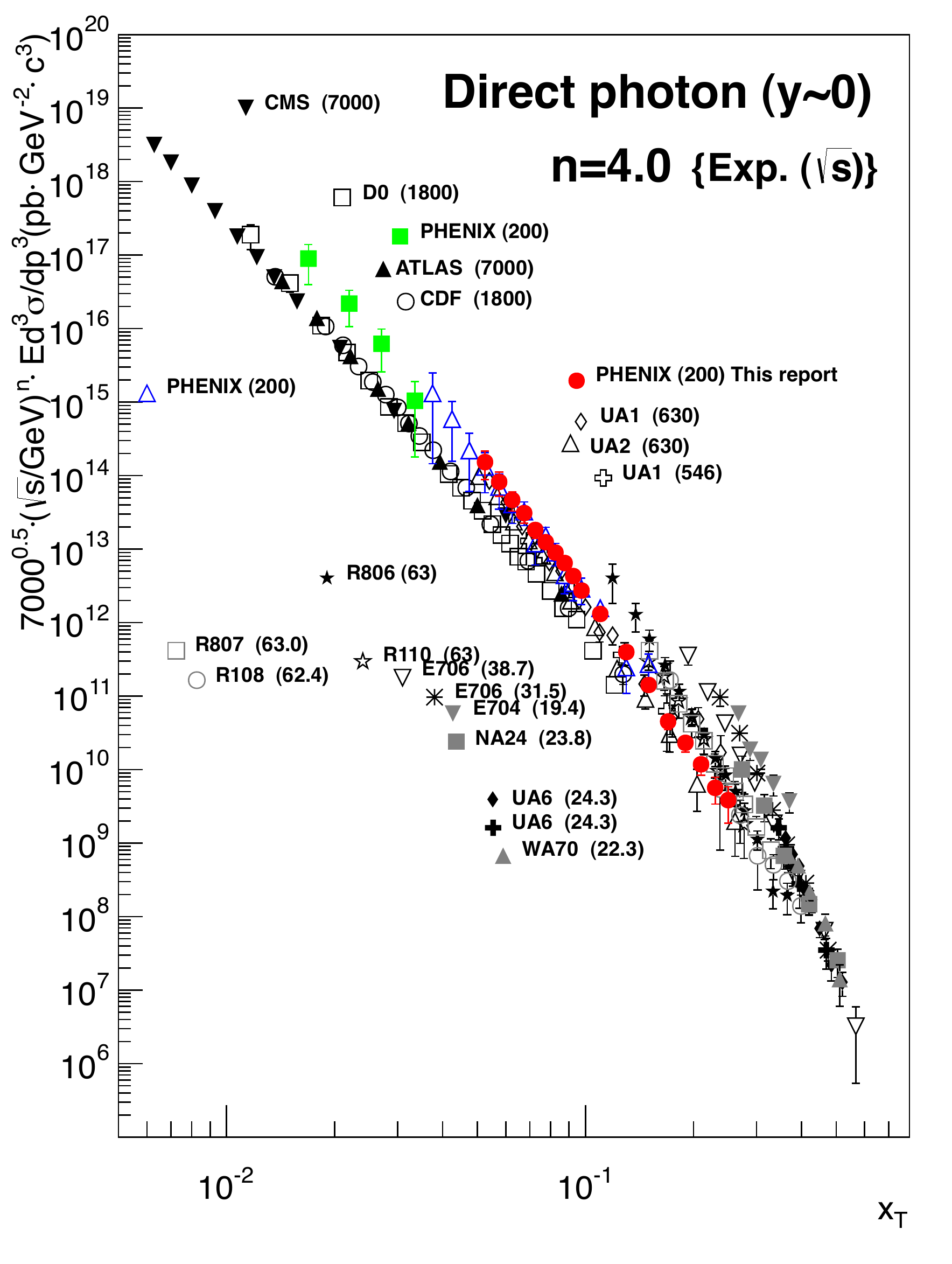}}
\end{center}\vspace*{-2pc}
\caption[]{a)(left) $\sqrt{s}^{\,n_{\rm eff}} \times E d^3\sigma/dp^3$, as a function of $x_T=2p_T/\sqrt{s}$, with $n_{\rm eff}=4.5$,  for direct-$\gamma$ measurements in p-p and $\bar{\rm p}$-p experiments at the ($\sqrt{s}$ GeV) indicated.~\cite{ppg136}. b)(right) same as (a) with $n_{\rm eff}=4.0$  
\label{fig:ggg}}\vspace*{-1pc}
\end{figure}

My favorite direct-$\gamma$ result this year is the improved PHENIX measurement in p-p collisions at $\sqrt{s}=200$ GeV out to $p_T=25$ GeV/c~\cite{ppg136}, in excellent agreement with pQCD. A more direct way to show this without a detailed theory calculation is to use $x_T$ scaling. Figure~\ref{fig:ggg}a shows $x_T$ scaling for all presently existing direct-$\gamma$ data\footnote{This includes the PHENIX p-p direct-virtual-$\gamma$ measurement down to $p_T\approx1$ GeV/c, further confirming the absence of a soft production mechanism for direct-$\gamma$ in p-p collisions~\cite{EriceProcPR}.}, with $n_{\rm eff}=4.5$, very close to the pure-scaling parton-parton Rutherford scattering result of $n_{\rm eff}=4.0$ (Fig.~\ref{fig:ggg}b). The deviation of the data points in Fig.~\ref{fig:ggg}b from the universal curve for $\sqrt{s}>38.7$ GeV is an illustration of the non-scaling of the coupling constant, structure and fragmentation functions  in \QCD---what I like to call ``\QCD\ in action". For $\sqrt{s}\leq 38.7$ GeV, the deviation of the data from the universal curve in Fig.~\ref{fig:ggg}a (and from pQCD) is claimed to be due to the $k_T$ effect (transverse momentum of the quarks in a nucleon).

Since the $p^{\gamma}_{T}$ of a direct-$\gamma$ can be measured very precisely, the fragmentation function of the jet from the away quark in the reaction $g+q\rightarrow \gamma +q$ can be measured by the direct-$\gamma-h$ correlations (where $h$ represents charged hadrons opposite in azimuth to the direct-$\gamma$) because the  $p_T$ of the away-quark at production is equal and opposite to $p^{\gamma}_{T}$, thus known to high precision (modulo a small $k_T$-smearing effect). 
\begin{figure}[hbt]
\begin{center}
\begin{tabular}{cc}
\hspace*{-0.15pc}\includegraphics[width=0.49\textwidth]{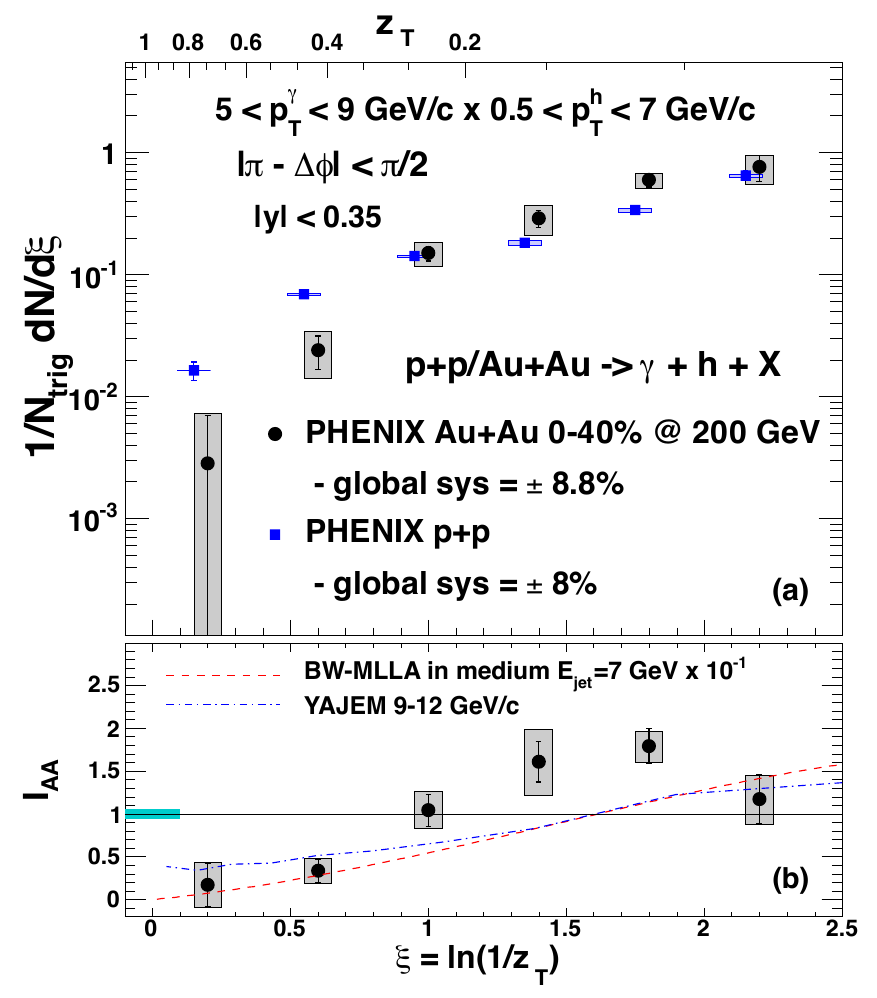}
\hspace*{-0.15pc}\includegraphics[width=0.49\textwidth,height=0.49\textwidth]{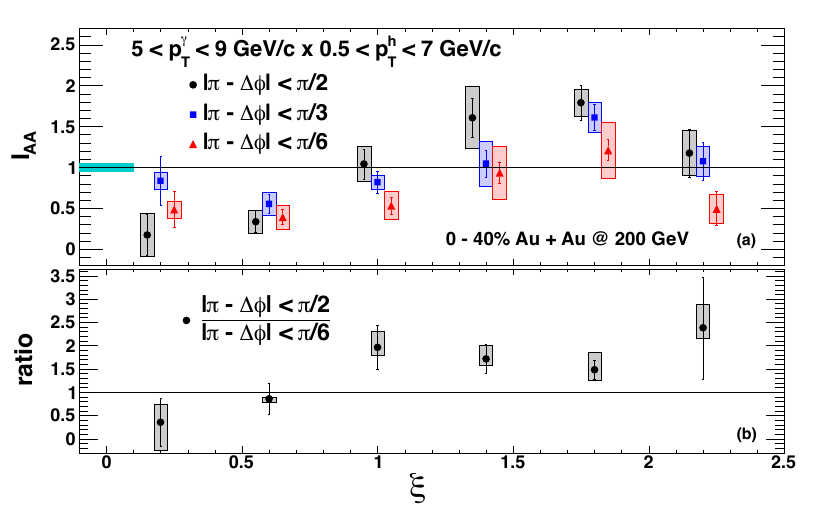}
\end{tabular}

\end{center}
\vspace*{-2pc}
\caption[]
{a)$\xi=-\ln (p^{h}_{T}/p^{\gamma}_{T})=-\ln{z_T}$ distributions of hadrons opposite in azimuth ($|\pi-\Delta\phi|<\pi/2$) to a direct-$\gamma$ trigger with $5<p^{\gamma}_{T}<9$ GeV/c in p-p and $0-40\%$ central Au+Au collisions at $\sqrt{s_{NN}}=200$ GeV from  PHENIX~\cite{JustinQM2012}.  b) Ratio of the two distributions, $I_{AA}(\xi)$. c) (right)-(top) $I_{AA}(\xi)$ as in (b) when the away-side azimuthal range is restricted as indicated. d) (right)-(bottom) Ratio of $I_{AA}$ for $|\pi-\Delta\phi|<\pi/2$ to $|\pi-\Delta\phi|<\pi/6$. 
\label{fig:PXgamma-h} }\vspace*{-1.5pc}
\end{figure}
This year, improved measurements by PHENIX~\cite{JustinQM2012} in both p-p and Au+Au collisions (Fig.~\ref{fig:PXgamma-h}) now indicate a significant modification of the fragmentation function in Au+Au (0-40\%) central collisions compared to p-p (Fig.~\ref{fig:PXgamma-h}a), with an enhancement at large $\xi=-\ln z_T$ (low $z_T=p^{h}_{T}/p^{\gamma}_{T}$) and a suppression at small $\xi$ (large $z_T$) which is more clearly seen as $I_{AA}(\xi)$, the ratio of the fragmentation functions in Au+Au/pp (Fig.~\ref{fig:PXgamma-h}b). As shown in Fig.~\ref{fig:PXgamma-h}c,d, restricting the away-side azimuthal range reduces the large $\xi>0.9$ ($p^{h}_{T}\lsim 3$ GeV/c) enhancement but leaves the suppression at small $\xi<0.9$  relatively unchanged, which shows that the large $\xi$ enhancement is predominantly at large angles, similar to the effect observed by CMS with actual jets.~\cite{CMSQM2012}. 

\begin{figure}[!h]
\begin{center}
\includegraphics[width=0.45\textwidth]{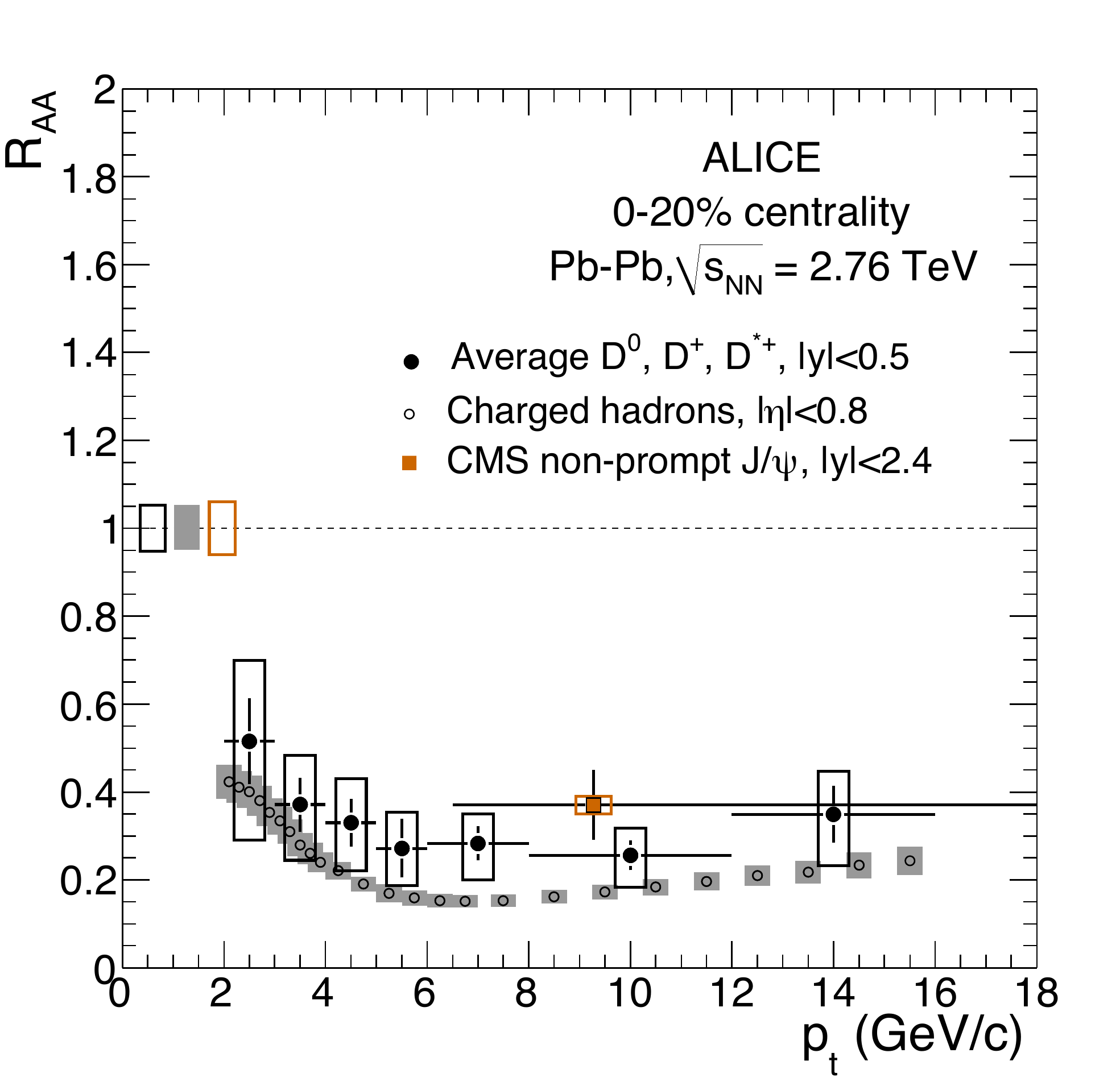}
\includegraphics[width=0.54\textwidth]{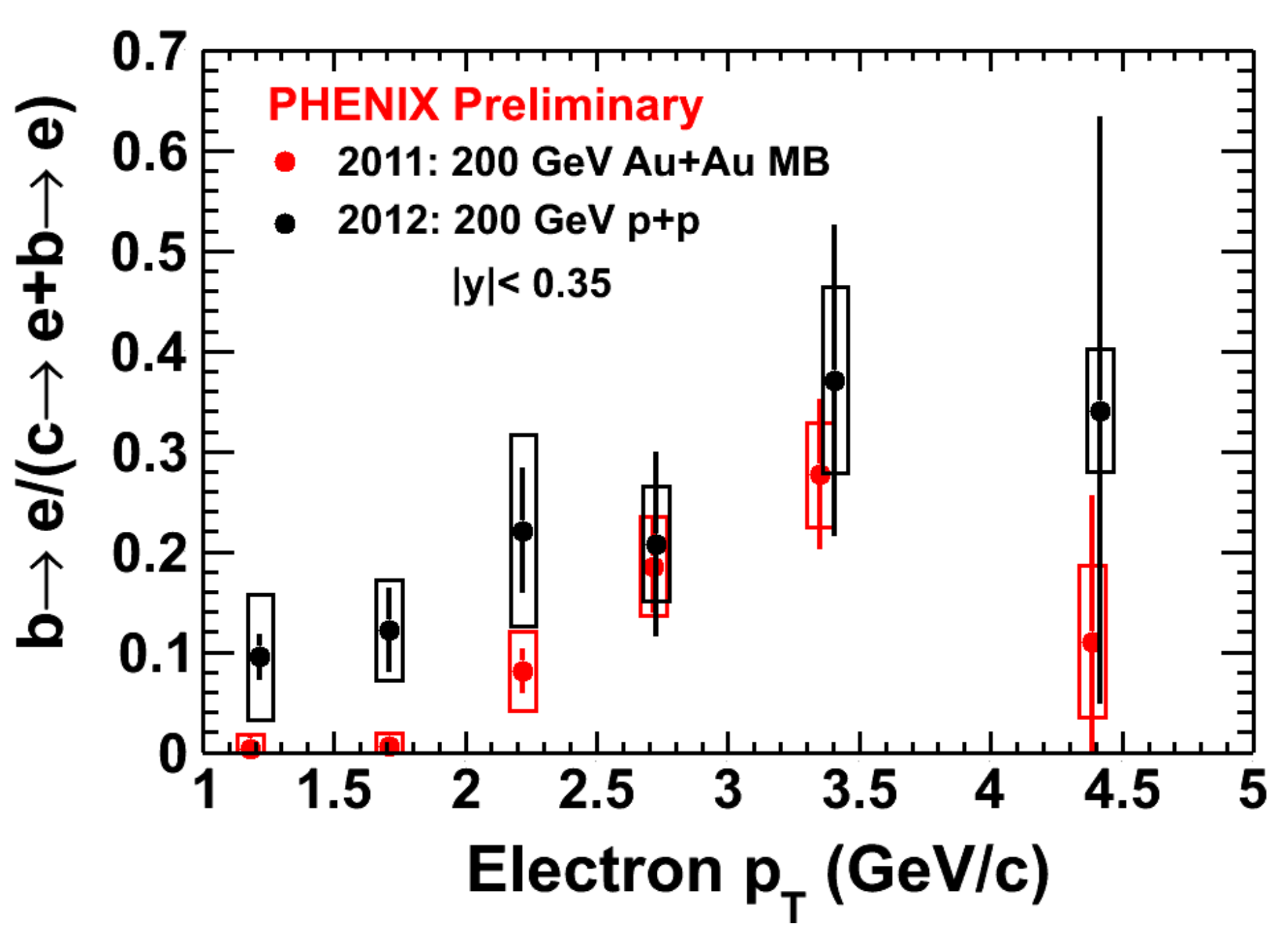}
\end{center}\vspace*{-1.5pc}
\caption[]{a) (left) $R_{AA}$ of ALICE~\cite{ALICEDsuppression} $D$-mesons, charged hadrons, and CMS~\cite{CMSUps1Ssupp} non-prompt $J/\Psi$, in central (0-20\%) Pb+Pb collisions at $\sqrt{s_{NN}}=2.76$ TeV b) (right) $b$-quark fraction ${F}_{b}={b\rightarrow e}/(c\rightarrow e+b\rightarrow e)$  of direct single-$e^{\pm}$ in p-p and Au+Au from PHENIX measurement of the Distance of Closest Approach (DCA) of the displaced vertex. }
\label{fig:heavy}\vspace*{-0.5pc}
\end{figure}

One of the most exciting discoveries at RHIC, now confirmed at the LHC, is the suppression of heavy quarks by the same amount as light quarks for $p_T\gsim 5$ GeV/c as indicated at RHIC (Fig.~\ref{fig:Tshirt}b) by direct single-$e^{\pm}$ from heavy quark ($c$, $b$) decay;   and at the LHC by $D$ mesons from $c$-quarks~\cite{ALICEDsuppression}, and non-prompt $J/\Psi$ from $b$-quarks~\cite{CMSUps1Ssupp} (Fig.~\ref{fig:heavy}a). 
The discovery at RHIC was a total surprise and a problem since it appears to disfavor the radiative energy loss explanation~\cite{BDMPS} of suppression (also called jet-quenching) because heavy quarks should radiate much less than light quarks or gluons. 

Many explanations have been offered including some from string theory; but the explanation I prefer was by Nino Zichichi~\cite{AZYukawa} who proposed that since the standard model Higgs Boson,  which gives mass to the Electro-Weak vector Bosons, does not necessarily give mass to Fermions, ``it cannot be excluded that in a QCD coloured world (a \QGP), the six quarks are all nearly massless''. If this were true it would certainly explain why light and heavy quarks appear to exhibit the same radiative energy loss in the medium. This idea can, in fact, be tested because the energy loss of one hard-scattered parton relative to its partner, e.g. $g+g\rightarrow b+\bar{b}$ , can be measured by experiments at RHIC and LHC using two particle correlations in which both the outgoing $b$ and $\bar{b}$ are identified by measurement of the Distance of Closest Approach (DCA) of their displaced decay vertices in silicon vertex detectors. When such results are available, they can be compared to $\pi^0$-charged hadron correlations from light quark and gluon jets, for which measurement of the relative energy loss has been demonstrated at RHIC~\cite{EriceProcPR}. 

Of course, measurement of the Yukawa couplings to Fermions of the candidate 125 GeV Higgs Boson at the LHC may be available by the end of 2012; but, already this year, the first direct measurement of $b$-quarks in p-p and Au+Au collisions at RHIC by their displaced vertices was made in the new PHENIX Silicon VTX detector~\cite{RachidQM2012}.  Figure~\ref{fig:heavy}b shows the measurement of the $b$-quark fraction ${F}_{b}={b\rightarrow e}/(c\rightarrow e+b\rightarrow e)$ of direct single $e^{\pm}$ in p-p and Au+Au collisions at $\sqrt{s_{NN}}=200$ GeV, using the PYTHIA $c$ and $b$ quark $p_T$ distributions in p-p collisions to calculate the DCA distributions of the $e^{\pm}$ in both p-p and Au+Au.  The fact that the Au+Au measurements for all $p_T^e$ are below the p-p measurements indicates clearly that the $b$-quark $p_T$ spectrum is modified in Au+Au compared to p-p. However, the correct conditional DCA distribution requires the actual (modified) $b$-quark $p_T$ spectrum in Au+Au,  which must be obtained by iteration. Once the iteration has converged, the $R^b_{AA}(p_T^e)$ can be calculated from the measured $R_{AA}(p_T^e)$ of the direct-single-$e^{\pm}$ by the relation $R^b_{AA}(p_T^e)=R_{AA}(p_T^e)\times {F}_{b}^{AA}/{F}_{b}^{pp}$. For example, if the final ${F}_{b}^{AA}={F}_{b}^{pp}$ then $R^b_{AA}(p_T^e)=R_{AA}(p_T^e)$. 


\end{document}